\documentclass[journal, trans]{IEEEtran}
\usepackage{amsmath,epsfig}
\usepackage{algorithm}
\usepackage{algpseudocode}
\usepackage{booktabs}
\usepackage[numbers,sort&compress]{natbib}
\usepackage{array}
\usepackage{subfigure}
\usepackage{amssymb}
\usepackage{color}
\ifCLASSINFOpdf
\else
\fi
\begin{document}
%
\title{Optimal Clustering Framework for Hyperspectral Band Selection}
%
%

\author{Qi~Wang,~\IEEEmembership{Senior Member,~IEEE,}
        Fahong~Zhang,
        and~Xuelong~Li,~\IEEEmembership{Fellow,~IEEE}
\thanks{This work was supported by the National Key R\&D Program of China under Grant 2017YFB1002202, National Natural Science Foundation of China under Grant 61773316, Fundamental Research Funds for the Central Universities under Grant 3102017AX010, and the Open Research Fund of Key Laboratory of Spectral Imaging Technology, Chinese Academy of Sciences.}

\thanks{Q. Wang is with the School of Computer Science, with the Center for Optical Imagery Analysis and Learning (OPTIMAL)
and with the Unmanned System Research Institute (USRI), Northwestern Polytechnical University,
Xi'an 710072, Shaanxi, China. E-mail: crabwq@gmail.com.}
\thanks{F. Zhang is with the School of Computer Science and the Center for Optical Imagery Analysis and Learning (OPTIMAL),
Northwestern Polytechnical University, Xi'an 710072, Shaanxi, China. E-mail: zfh@mail.nwpu.edu.cn. }
\thanks{X. Li is with the Xi'an Institute of Optics and Precision Mechanics, Chinese Academy of Sciences, Xi'an 710119, Shaanxi, P. R. China and with the University of Chinese Academy of Sciences, Beijing 100049, P. R. China. Email: xuelong\_li@opt.ac.cn.
}
\thanks{\copyright 2018 IEEE. Personal use of this material is permitted. Permission from IEEE must be obtained for all other uses, in any current or future media, including reprinting/republishing this material for advertising or promotional purposes, creating new collective works, for resale or redistribution to servers or lists, or reuse of any copyrighted component of this work in other works.
}
\thanks{Q. Wang, F. Zhang, and X. Li, ``Optimal clustering framework for hyperspectral band selection,'' IEEE Trans. Geoscience and Remote Sensing, vol. 56, no. 10, pp. 5910-5922, 2018, 10.1109/TGRS.2018.2828161.
}
}

\maketitle

\begin{abstract}
Band selection, by choosing a set of representative bands in hyperspectral image (HSI),
is an effective method to reduce the redundant information without compromising the original contents.
Recently, various unsupervised band selection methods have been proposed, but most of them are based on approximation algorithms which can only obtain suboptimal solutions toward a specific objective function.
This paper focuses on clustering-based band selection, and proposes a new framework to solve the above dilemma, claiming the following contributions:
1) An optimal clustering framework (OCF),
which can obtain the optimal clustering result for a particular form of objective function under a reasonable constraint.
2) A rank on clusters strategy (RCS), which provides an effective criterion to select bands on existing clustering structure.
3) An automatic method to determine the number of  the required bands,
which can better evaluate the distinctive information produced by certain number of bands.
In experiments, the proposed algorithm is compared to some state-of-the-art competitors.
According to the experimental results, the proposed algorithm is robust and significantly outperform the other methods on various data sets.
\end{abstract}


%
\IEEEpeerreviewmaketitle

\section{Introduction}
\label{sec:introduction}
%
%
%
%
\IEEEPARstart{H}{yperspectral} image (HSI) processing has attracted considerable attention in recent years.
HSIs can provide rich band information from different wavelengths and thus get widely used in various research field,
such as biological analysis \cite{DBLP:journals/tgrs/LuoYCZ13} and medical imaging processing \cite{DBLP:journals/tbe/AkbariKKT10}.
HSIs record the reflectances of electromagnetic waves of different wavelengths,
and the reflectance of each electromagnetic wave is stored in a 2-D image.
Hence, a HSI is a data cube which contains hundreds of 2-D images.
Though significant success in the field of HSI application has been obtained,
how to deal with the large dimensional data is still a challenging problem,
since high correlations and dependencies among them cause huge computational complexity as well as ``Hughes phenomenon'' \cite{DBLP:journals/tit/Hughes68}.
In view of this, the reduction in HSI is deemed to be a very important work.

Generally, HSI reduction can be achieved by feature extraction or feature selection (also known as band selection) techniques.
For feature extraction \cite{7530874, 8116758, 7896571, Kang2014Feature, Kang2014Intrinsic}, the original HSI is projected into a lower dimensional space and a reduced data set is generated.
While for band selection, some discriminative bands are chosen to represent the original data set without modification.
In experiments,
feature extraction could usually achieve better performance.
However, band selection is usually more preferred
since it can can preserve the information of the original data in physical sense,
making the reduced data sets more interpretable.

According to the involvement of the labeled and the unlabeled samples,
band selection can be divided into supervised \cite{7378270, 7729714, 7807260, 5530350, 8059826},
semi-supervised \cite{6977960, 6738646, 6738664, 7332945} and unsupervised \cite{6853323, 7888954, 7214263, DBLP:journals/tnn/WangLY16, 7378877, 7166333} methods.
Supervised and semi-supervised methods utilize the labeled samples to guide the selection process.
However, since the acquisition of the labeled samples is a difficult task, sometimes they are not very practical in real application.
Therefore, we mainly focus on unsupervised band selection in this paper.

According to the employed searching strategy \cite{DBLP:journals/tip/YuanZL17},
unsupervised band selection can be categorized into ranking-based, clustering-based, greedy-based and evolutionary-based methods.
Ranking-based methods assign each band a rank value and simply select the top-rank bands with the desired number.
Clustering-based methods first separate all the bands into clusters,
and then select the most representative bands in each cluster to constitute the band subset.
Greedy-based methods are iterative processes.
In each iteration, the currently optimal band will be selected founded on the previous results.

As for evolutionary-based methods,
they first generate a candidate band set,
and then repeatedly update it via a specific evolutionary strategy
until the convergence condition is satisfied.

Taking an overall review of the above mentioned kinds of band selection methods,
one issue can be found that the existing methods can only obtain an approximately optimal solution.
For example,
greedy-based methods are only optimal in current iteration, rather than in global.
Also, evolutionary-based methods are based on some random processes and usually trapped into local optimums.
The main reason of this phenomenon is that the solution space of band selection problem is too large to attain the optimal solution in limited time.
For point-wise selection \cite{7807260}, the number of ways to select $K$ bands from a $L$ bands HSI is $\binom{L}{K}$.
For group-wise selection \cite{7807260}, the number of ways to cluster the bands increases to the Stirling Number of the Second Kind \cite{957546},
denoted as $\left\{\begin{smallmatrix}L \cr K\end{smallmatrix}\right\}$.
Suppose $L=200$ and $K=15$, the value of $\binom{L}{K}$ is about $10^{22}$,
and the value of $\left\{\begin{smallmatrix}L \cr K\end{smallmatrix}\right\}$ is about $10^{223}$.
Consequently,
achieving an optimal solution is considered to be a very difficult task, especially for clustering-based methods.

In this paper, we focus on clustering-based methods,
and propose a general framework which can uncover the optimal clustering structure under a reasonable constraint.

The main contributions of this paper are listed as follows.

1) An \emph{optimal clustering framework} (OCF) is proposed to search for the optimal clustering structure in HSI.
Though achieving the optimal clustering result has been demonstrated NP-hard for many kinds of objective function,
the proposed OCF can still find the optimal solution under a reasonable constraint for HSI data sets.
Moreover, the proposed OCF is a general framework,
which means different kinds of objective function can be optimized via the same procedure
once they comply with the specific form.

2) A \emph{ranking on clusters strategy} (RCS) is proposed as an effective criterion to select the representative bands under the achieved clustering structure.
By applying an arbitrary ranking algorithm on the clustering result,
RCS can better exploit the advantages of clustering-based and ranking-based methods,
and generate a band subset with lower correlation and more discriminative information.

3) An automatic method to determine the required number of bands is proposed.
Through reducing the correlation among bands, we aim to uncover how much distinctive information
can be produced by certain number of bands.
Experiments show that this method can offer a promising estimation of band number for various data sets.

The remainder of this paper is organized as follows.
In Section \ref{sec:related_work}, related unsupervised band selection methods are introduced.
In Section \ref{sec:framework}, OCF and RCS are formulated.
Then in Section \ref{sec:Implementation}, we show some instances to implement a band selection algorithm based on OCF and RCS.
After that, the experimental results on four real HSI data sets are shown in Section \ref{sec:experiment}.
Finally, conclusions are made in Section \ref{sec:conclusion}.
\begin{figure*}[htb]
\begin{minipage}[b]{1.0\linewidth}
  \centering
  \centerline{\epsfig{figure=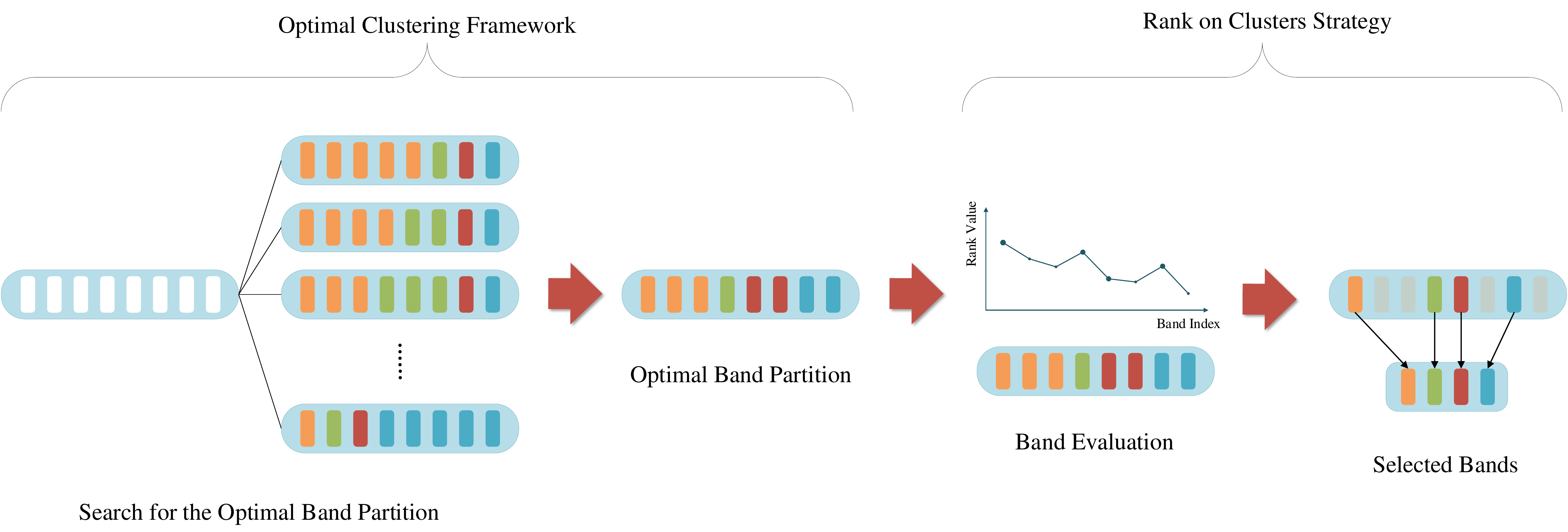,width=17cm}}
\end{minipage}
\caption{The flowchart of the overall procedure to conduct band selection.
First, given an objective function, we search for the optimal band partition via OCF (in the figure, each color refers to a cluster of bands).
Then all the bands are evaluated through an arbitrary ranking method.
Finally the top-rank bands are selected in each cluster.
}
\label{fig:flowchart}
\end{figure*}
\section{Related Work}
\label{sec:related_work}
As introduced in Section \ref{sec:introduction}, unsupervised band selection can be categorized into
ranking-based, clustering-based, greedy-based and evolutionary-based methods.
In this section, some representative band selection methods will be introduced sequentially.

\subsection{Ranking-based methods}
Ranking-based methods \cite{DBLP:journals/tgrs/ChangDSA99, DBLP:journals/tgrs/ChangW06} aim at designing a criterion to
evaluate the importance of each band, and use the top-rank bands to constitute the band subset.
The advantage of ranking-based methods is that the most discriminative bands can be discovered.
However, there is usually high correlation among the selected bands.
To give examples, some representative ranking-based methods will be introduced.

Maximum-variance principal component analysis (MVPCA) \cite{DBLP:journals/tgrs/ChangDSA99}
is a joint band prioritization and band-decorrelation approach.
In MVPCA, a data-sample covariance matrix is firstly constructed.
Second, eigenvalue decomposition is performed over the covariance matrix,
and a loading factor matrix is constructed.
Finally, all the bands are prioritized according to the loading factor matrix.
From another point of view, bands are prioritized by their variances essentially.
MVPCA is rational since bands with higher variances contain more distinct information of ground objects and are more discriminative in general.
Nevertheless, it is sensitive to noisy bands since they usually have large variances.
Moreover, the top-rank bands are usually highly correlated,
leading to a large amount of redundant information among the selected bands.

Constrained band selection (CBS) \cite{DBLP:journals/tgrs/ChangW06} is a band correlation minimization process.
It first design a finite impulse response (FIR) filter for each band,
and minimize the averaged least squares filter output.
Then bands are ranked according to the solution of the above minimization problem since
it can measure the correlation between one particular band and the entire HSI.
CBS is less sensitive to noisy bands because they usually have lower correlation with the other bands,
and hence will be assigned with smaller ranking values.
But similar to MVPCA, the top-rank bands may still be highly correlated since the interaction among the selected bands are neglected.

\subsection{Clustering-based methods}
Clustering-based methods \cite{DBLP:journals/tgrs/UsoPSG07, 7161371, 7729628, 7295589, 6165389, su2012hyperspectral} first separate the whole bands into clusters,
and then select one in each to constitute the band subset.
Unlike ranking-based methods,
clustering-based methods focus on the reduction of the correlation among bands.
In the following, several representative methods are listed and discussed.

In WaLuMI and WaLuDi \cite{DBLP:journals/tgrs/UsoPSG07},
hierarchical clustering is utilized to partition the bands into clusters.
In order to measure the distances among these bands,
two criteria are involved,
known as mutual information and Kullback-Leibler divergence.
First, the total band set is separated into clusters using the Ward's linkage method.
After that, the band which has the highest similarity with the other bands is selected in each cluster.
Through hierarchical clustering, WaLuMI and WaLuDi can effectively reduce the correlation among bands.
However, these methods are sensitive to noisy bands since they usually have low correlations with the others and easy to form single-band clusters.

Enhanced fast density-peak-based clustering (E-FDPC) \cite{7161371} can be viewed as a clustering-based as well as a ranking-based method.
Based on the idea that a cluster center should have large local density and inter-cluster distance,
E-FDPC prioritizes each band via combining these two indicators.
Then similar to ranking-based methods, the top-rank bands are selected to constitute the band subset.
Through investigating the intra-cluster distance of bands,
E-FDPC can reduce the possibility that two correlated bands are selected simultaneously.
Nevertheless, it is difficult to measure the local density and intra-cluster distance exactly.

Squaring weighted low-rank subspace clustering (SWLRSC) \cite{7729628} firstly constructs a low-rank coefficient matrix by solving the low-rank optimization problem.
Then a squaring weighted strategy is used to obtain the similarity matrix.
After that, the spectral clustering is conducted using the matrix and the clustering result is generated.
Finally, the bands which are closest to the centroid of each cluster are chosen to constitute the band subset.
By constructing such a low-rank coefficient matrix,
the relationship among bands can be better exploited.
However, it's time consuming to find such a low-rank presentation for large-scale HSI data sets.
\subsection{Greedy-based methods}
Greedy-based methods \cite{DBLP:journals/tgrs/GengSJZ14, 6783712, DBLP:journals/tnn/WangLY16}
greedily select or remove one band from the candidate band subset,
and make sure that the objective function is optimized in each iteration.
Though the optimal solution cannot be obtained,
greedy-based methods still offer a good substitution to it.
In the following, two representative methods are introduced.

Volume gradient band selection (VGBS) \cite{DBLP:journals/tgrs/GengSJZ14} is a geometry-based method,
in which bands are treated as points lying in a high dimensional space.
In the beginning of the algorithm, all of the bands are considered as candidate bands and constitute a parallelotope.
Then, bands lead to the minimal losses of the parallelotope volume will be removed repeatedly.
The algorithm stops when the desired number of bands is remained in the subset.
VGBS can effectively reduce the correlation among bands
since low-correlated bands often construct a large-volume parallelotope.
Nevertheless, noisy bands often contribute a lot to this volume and are easy to be selected.
Hence VGBS often do not perform well in data sets with large noises.

Different from VGBS, sequential forward selection (SFS) \cite{6783712} starts with an empty band subset,
and iteratively adds bands to it until the desired number of them have been obtained.
Minimum estimated abundance covariance (MEAC) is the objective function in SFS.
In each iteration, the band which can maximally decrease the MEAC value of current band subset will be selected.
SFS method is efficient in computation,
but sensitive to the initial condition.

\subsection{Evolutionary-based methods}
Evolutionary-based methods \cite{DBLP:journals/tgrs/YuanZW15,7332945,7214263,7536149}
first generate a band subset with the desired number of bands randomly,
and then apply some evolutionary algorithms to update it, seeking a nearly optimal solution.
Here two representative methods are introduced.

In multi-task sparsity pursuit (MTSP) \cite{DBLP:journals/tgrs/YuanZW15},
a compressive band descriptor is first constructed as a reduction of the original HSI.
Then a multi-task learning based criterion is proposed to evaluate the effectiveness of the band descriptor.
Finally the immune clonal strategy is utilized to search for the optimal band combination.
Compared to greedy-based methods, the utilized immune clonal strategy is proved with powerful global searching ability.
However, in order to accelerate the calculation of the objective function,
some intrinsic information is lost when constructing the compressive descriptor.

Multi-objective optimization band selection (MOBS) \cite{7214263} presents a multi-objective model for band selection,
in which information entropy and the number bands are considered as two objective functions.
They are optimized simultaneously by a multi-objective evolutionary algorithm.
MOBS is less sensitive to parameters and can obtain a more stable performance,
but using the sum of information entropy to evaluate a band subset
is sometimes too simple to capture the interrelationship among bands.

\section{Optimal Clustering Framework}
\label{sec:framework}
This section details the proposed OCF
and introduces the overall procedure as shown in Fig. \ref{fig:flowchart} to design a band selection algorithm.
First, the idea of dynamic programming is briefly reviewed as the background knowledge.
Second, the rationality of CBIC is analysed and discussed.
Third, the proposed framework which can generate the optimal clustering result is introduced and demonstrated.
After that, a more general framework is shown as an extension to the original one.
Finally, we propose a novel strategy to determine the selected bands based on the achieved clustering result.

\subsection{Introduction to Dynamic Programming}
\emph{Dynamic programming} (DP) is an effective optimization technique that was proposed by Berman (R.Bellman), etc. in 1951 \cite{DBLP:journals/tssc/JefferisF65, 1098129}.
In DP, a complex problem is repeatedly broken down into a series of subproblems until they are simple enough to be resolved directly.
Then these subproblems are constantly combined to solve the more complex ones and finally solve the original problem.
One problem must have two attributes to be suitable for DP.

1) Optimal substructure. This attribute says
a problem can be broken down into simpler subproblems with the same form,
and the solution of the original problem can be obtained through these subproblems.

2) Overlapping subproblems. This means that while some subproblems are broken down into simpler ones,
there should be overlaps among them.
In other words, a subproblem should be reused several times for the solving of different and more complex subproblems.
If this attribute is not satisfied,
the number of subproblems will increase exponentially,
bringing unaffordable computational complexity.

\subsection{Contiguous Band Indexes Constraint}
\label{sec:CBIC}
\begin{figure}[htb]

\begin{minipage}[b]{1.0\linewidth}
  \centering
  \centerline{\epsfig{figure=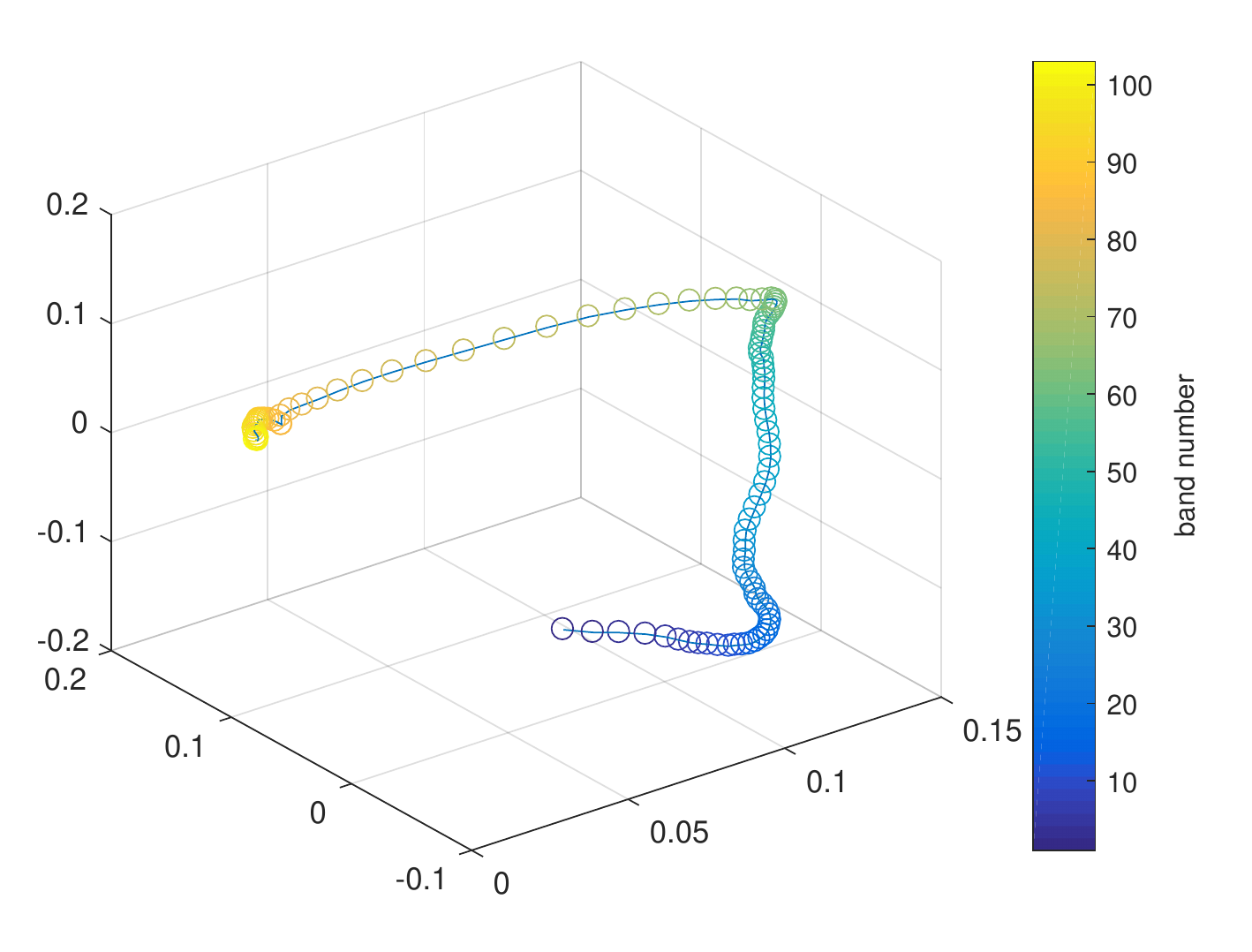,width=7cm}}
\end{minipage}
\caption{Spectral bands projected to a three-dimensional feature space through PCA transformation on Pavia University data set.}
\label{fig:CBIC}
\end{figure}

In HSI, bands represent the reflectance of the scene to electromagnetic waves in various wavelengths.
Seen from physical view, electromagnetic waves with closer wavelengths produce similar reflectances.
Hence bands with similar wavelengths usually have stronger correlation \cite{Zhang2018Cluster, Wei2017Structured}.
Fig. \ref{fig:CBIC} plots all the bands in Pavia University data set,
in which each band is vectorized and projected to a three-dimensional feature space through PCA transformation.
As we can see, when we connect the bands in the order of their indexes, we get a very smooth curve.
This phenomenon further proves that strong correlation exists between bands with close wavelengths.
Inspired by this, we propose a \emph{contiguous band indexes constraint} (CBIC) for band clustering problem,
which claims that bands in the same cluster should have contiguous wavelengths.

According to the analysis in Section \ref{sec:introduction}, the number of all the possible solutions for band clustering problem is $\left\{\begin{smallmatrix}L \cr K\end{smallmatrix}\right\}$.
Under the proposed constraint, this number is reduced to $\binom{L-1}{K-1}$ .
When $L=200$ and $K=15$, the value of $\binom{L-1}{K-1}$ is about $10^{21}$,
which is much smaller than that of $\left\{\begin{smallmatrix}L \cr K\end{smallmatrix}\right\}$ ($\approx 10^{223}$).
Consequently, the proposed constraint can effectively reduce the size of the solution space for band clustering problem.

Owing to CBIC,
the original clustering problem is converted to
finding a series of critical bands to separate the whole bands into intervals.
This enables us to search for the optimal solution in a more efficient way.
\subsection{OCF Formulization}
\label{ssec:problem_formulization}
This subsection details the proposed \emph{optimal clustering framework} (OCF).
First, we define some notations that will be used throughout the paper.
Denote $\mathbf{x}_{l} \in \mathbb{R}^{N \times 1}$ as the $l$-th band vector
and $X_{i}^{j} = \left\{\mathbf{x}_{l}\right\}_{l = i}^{j}$ for each $1\leq i\leq j\leq L$ as band intervals.
Here $N$ is the number of pixels, and $L$ is the number of bands in HSI.
$\mathbf{s} = (s_{1}, ..., s_{K-1})^{T}$ specifies the \emph{critical band indexes vector} (CBIV), in which $0 < s_{1} <... < s_{K-1} < L$.
Here $s_{i}$ is the index of the $i$-th critical band,
i.e., the last band of cluster $i$, and $K$ is the number of selected bands.
For convenience of expression, we set $s_0=0,s_K=L$.
Noting that once $\mathbf{s}$ is determined, the whole band set can be separated into $K$ subsets:
$X_{s_{0}+1}^{s_{1}}$, $X_{s_{1}+1}^{s_{2}}$,..., $X_{s_{K-1}+1}^{s_{K}}$.

In a clustering algorithm, how to design an effective objective function is an important issue.
A straightforward idea is to individually evaluate the contribution of each cluster,
and then sum up these contributions as a measurement of the whole clustering result.
Though very simple,
this strategy has been adopted by many well-known clustering methods (e.g., k-means and spectral clustering \cite{1238361}) and demonstrated to be effective.

Following this idea, we define a mapping $f: \{X_i^j \mid 1 \leq i \leq j \leq L\} \rightarrow \mathbb{R} $ to evaluate the contribution of each band interval $X_i^j$.
Since the clustering result is determined by a series of critical bands under CBIC,
the contribution of cluster $i$ can be represented by $f(X_{s_{i-1}+1}^{s_{i}})$.
Based on the above consideration,
a general form of the objective function $D_s$ is given as:
\begin{align}
\label{Eq:objective_function}
    &D_s(\mathbf{s}) = \sum_{i = 1}^{K}f(X_{s_{i-1}+1}^{s_{i}}).
\end{align}
Without loss of generality, we assume that function $D_s$ is supposed to be maximized.
So our optimization problem turns to be:
\begin{equation}
\label{Eq:optimization_problem}
    \mathop{\max}\limits_{0 < s_{1} < ... < s_{K-1}<L}\sum_{i = 1}^{K}f(X_{s_{i-1}+1}^{s_{i}}).
\end{equation}

After the optimization problem is clarified,
the solution will be given in two steps,
named as problem decomposition and subproblem combination, respectively.
It should be pointed out that, the mapping $f$ here is still a general form,
which means the solution will be available for arbitrary definition of $f$.

1) Problem decomposition.
Let $M_l^k$ be the solution of a simpler subproblem of Eq. (\ref{Eq:optimization_problem}):
\begin{equation}
\label{Eq:optimization_problem2}
    M_l^k  = \mathop{\max}\limits_{0<s_1<...<s_k = l}\sum_{i = 1}^{k}f(X_{s_{i-1}+1}^{s_{i}}),
\end{equation}
where  $l \leq L$ and $k \leq \min(K, l)$.
Intuitively, $M_l^k$ is the optimal solution when our target is to partition the first $l$ bands into $k$ intervals
(the original one is to partition the whole $L$ bands into $K$ intervals).
Specially, $M_L^K$ is the solution of Eq. (\ref{Eq:optimization_problem}).

\begin{figure}[htb]
\begin{minipage}[b]{1.0\linewidth}
  \centering
  \centerline{\epsfig{figure=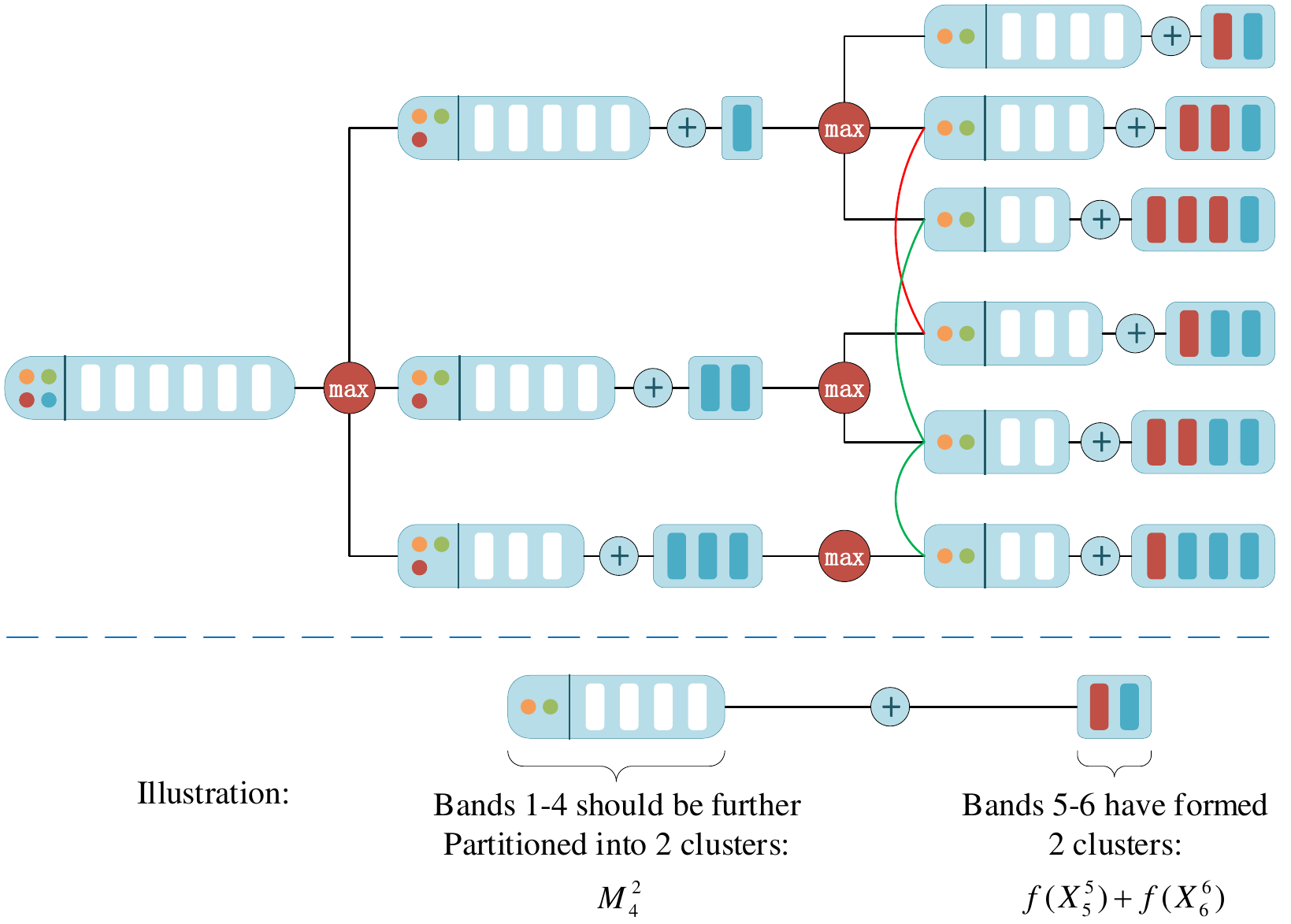,width=8.5cm}}
\end{minipage}
\caption{An example to calculate $M_6^4$.
As is illustrated, $M_6^4$ can be obtained by calculating the maximal value among $M_5^3+f(X_6^6)$, $M_4^3+f(X_5^6)$ and $M_3^3+f(X_4^6)$,
where $M_5^3$, $M_4^3$ and $M_3^3$ can be further decomposed into simpler subproblems.
It is worth noting that there are overlapped subproblems in the decomposition (connected by line with red and green color).
}
\label{fig:OCF_example}
\end{figure}

Then by enumerating all the possible value of $s_{k-1}$,
Eq. (\ref{Eq:optimization_problem2}) can be derived into:

\begin{equation}
\label{Eq:optimization_problem4}
    M_l^k = \mathop{\max}\limits_{k-1 \leq s_{k-1} < s_{k} = l}M_{s_{k-1}}^{k-1} + f(X_{s_{k-1}+1}^{s_k}).
\end{equation}
By substituting $k=1$ into Eq. (\ref{Eq:optimization_problem2}), we have:
\begin{equation}
\label{Eq:optimization_problem5}
    M_l^1 = f(X_1^l).
\end{equation}
For better comprehension of Eq. (\ref{Eq:optimization_problem4}),
an example to decompose this problem is given in Fig. \ref{fig:OCF_example}.

2) Subproblem combination.
The above analysis uncovers the optimal substructure attribute implied in the original problem.
According to Eq. (\ref{Eq:optimization_problem4}) and Eq. (\ref{Eq:optimization_problem5}),
all the $M_l^2$, for each $2 \leq l \leq L$, can be obtained since $M_l^1$ is already known.
Similarly, all the $M_l^3, M_l^4,..., M_l^K$ can also be achieved based on the previous results.
Through this order of calculation,
all the values of $M_l^k$ including $M_L^K$ can be obtained efficiently.

We are now at a point to recover the CBIV $\mathbf{s}^* = (s_1^*, s_2^*,..., s_{K-1}^*)^T$ corresponding to $M_L^K$,
since it is needed to acquire the final clustering result.
Denote $Q_l^k$ as the maximizer corresponding to $M_l^k$:
\begin{equation}
\label{Eq:optimization_problem6}
    Q_l^k = \mathop{\arg\max}\limits_{k-1 \leq s_{k-1} < s_{k} = l}M_{s_{k-1}}^{k-1} + f(X_{s_{k-1}+1}^{s_k}).
\end{equation}
It is easy to see that there is:
\begin{equation}
\label{Eq:optimization_problem7}
    s_{k-1}^{*} = Q_{s_{k}^*}^{k}.
\end{equation}
Since $s_{K}^* = L$ is known,
$s_{K-1}^*, s_{K-2}^*,...,s_{1}^*$ can be recovered by substituting $k = K, K-1,...,1$ into Eq. (\ref{Eq:optimization_problem7}) sequentially.

For more details about the framework, refer to the pseudo code shown in Algorithm \ref{Alg:OCF}.

\begin{algorithm}
  \caption{OCF ($D_s$ is maximized)}
  \begin{algorithmic}[1]
    \Require Set of bands $X_1^L$, mapping $f$ and cluster number $K$.
    \For{$l \gets 1$ \textbf{to} $L$}
        \State $M_l^1 \gets f(X_1^l)$
        \State $Q_l^1 \gets 0$
    \EndFor
    \For{$k \gets 2$ \textbf{to} $K$}
        \For{$l \gets k$ \textbf{to} $L$}
        \State $M_l^k \gets -\infty$
        \State $p^{*} \gets 0$
            \For{$p \gets k-1$ \textbf{to} $l-1$}
                \If{$M_l^k < M_p^{k-1} + f(X_{p+1}^{l})$}
                    \State $M_l^k \gets M_p^{k-1} + f(X_{p+1}^{l})$
                    \State $p^{*} \gets p$
                \EndIf
            \EndFor
        \State $Q_l^k \gets p^{*}$.
        \EndFor
    \EndFor
    \State $s_K^* \gets L$
    \For{$k \gets K-1$ \textbf{to} $1$}
    \State $s_{k}^* \gets Q_{s_{k+1}^*}^{k+1}$
    \EndFor
    \Ensure CBIV corresponding to $M_L^K$: $\mathbf{s}^* = (s_{1}^*,..., s_{K-1}^*)^T$.
  \end{algorithmic}
  \label{Alg:OCF}
\end{algorithm}

\subsection{Extended-OCF}
\label{sec:E-OCF}
In this part, the proposed OCF will be extended to a more general form,
so it can be suitable for more kinds of objective function that may be more effective.
In Eq. (\ref{Eq:objective_function}),
the objective function is defined as the sum of the contribution of each band interval.
In fact, we can attain a more general form of the objective function
via replacing the original sum operator by an arbitrary binary operator $\oplus$ : $\mathbb{R}^2 \rightarrow \mathbb{R}$ as follows.
\begin{align}
\label{Eq:E_OCF}
D_e(\mathbf{s}) = f(X_{s_0+1}^{s_1}) \oplus f(X_{s_1+1}^{s_2}) \oplus...\oplus f(X_{s_{K-1}+1}^{s_K}).
\end{align}
In fact, the optimization of Eq. (\ref{Eq:E_OCF}) is nearly the same as Eq. (\ref{Eq:objective_function}).
The only difference is to replace ``$+$'' operator with ``$\oplus$'' in Eq. (\ref{Eq:optimization_problem4}).

Consequently, given a tuple $(\oplus, f)$,
an objective function can be designed by Eq. (\ref{Eq:E_OCF}).
Also, it can be readily solved via OCF.

\subsection{Rank on Clusters Strategy}
\label{sec:RCS}
After the bands have been separated into clusters,
conventional clustering-based band selection methods usually select one band in each cluster individually by applying some kinds of criteria,
e.g., to select which is closest to the centroid \cite{7729628}.
However, this kind of strategy is found ineffective in experiments
since the most discriminative bands in each cluster may not be discriminative with respect to the total bands.

To tackle this problem,
a simple but effective strategy is proposed to select bands under the achieved clustering structure.
The basic idea is to rank the bands according to their discriminations,
and select those with higher rank values,
while ensuring that there is only one band selected in each cluster.

Here we give a more formal description.
First, bands are assigned with rank values, denoted as $\mathbf{r} = (r_1,r_2,...,r_L)^T$,
where $r_l$ is the rank value of the $l$-th band.
Then we use $\mathbf{g} = (g_1,g_2,...,g_L)^T$ to record the index of the cluster that each band belongs to.
To be specific,
if $s_{i-1}^* < l \leq s_{i}^*$, there is $g_l = i$.
Suppose the indexes of $K$ selected bands form a vector $\mathbf{b} = (b_1,b_2,...b_K)^T$.
Our purpose is to solve the following optimization problem:
\begin{equation}
\label{Eq:RCS1}
    \mathop{\max}\limits_{\mathbf{b}} \sum_{i=1}^{K}r_{b_i} \quad \mathbf{s.t.} \quad \forall j\neq k \in [1,K], \quad g_{b_j} \neq g_{b_k}.
\end{equation}
Obviously, the solution is to select the band with the highest rank value in each cluster:
\begin{equation}
\label{Eq:RCS2}
    b_i^* = \mathop{\arg\max}\limits_{s_{i-1}^{*} < l \leq s_{i}^{*}} r_l.
\end{equation}

Intuitively, RCS first assigns rank values to the bands,
and bands with the highest rank values in each cluster are selected to constitute the desired band subset.
In this way, both the discrimination of bands and the correlation among bands are taken into account to acquire a more superior band subset.

\section{Implementation of OCF}
\label{sec:Implementation}
In the previous section,
we have learnt that how to implement a band selection algorithm based on OCF when given an objective function and a ranking method.
In this section, the selection of these two factors and some other issues in implementing the framework will be discussed.
First, the objective functions and ranking methods utilized in this paper will be introduced.
Then, the computational complexity to implement the proposed algorithm will be analysed.
Finally, a novel method to identify the number of the selected bands will be presented.

\subsection{Objective Function}
In this paper, two objective functions are adopted to carry out OCF,
namely \emph{normalized cut} (NC) and \emph{top-rank cut} (TRC).

\textbf{Normalized cut criterion}.
NC is an effective graph-theoretic criterion that first adopted in \emph{spectral clustering} (SC) \cite{DBLP:journals/pami/ShiM00, 1238361}.
Assume there is a weighted undirected graph: $G = (V, W)$,
where $V = \{1,2,...L\}$ is the node set and $W$ with entries $w_{ij}$ is the similarity matrix.
A $K$-way partition of $G$ can be denoted by $\Gamma_{V}^{K} = \{V_{1}, V_{2},..., V_{K}\}$,
in which $V = \cup_{i = 1}^{K}V_{i}$ and $\forall i \neq j, V_{i} \cap V_{j} = \varnothing$.
The NC and \emph{normalized association} (NA) \cite{1238361} are defined as:
\begin{align}
\label{Eq:NC}
NC(\Gamma_{V}^{K}) = \frac{1}{K}\sum_{i=1}^{K}\frac{\sum_{j\in V_{i}, k \notin V_{i}}w_{jk}}{\sum_{j\in V_{i}, k \in V}w_{jk}}, \\
\label{Eq:NA}
NA(\Gamma_{V}^{K}) = \frac{1}{K}\sum_{i=1}^{K}\frac{\sum_{j\in V_{i}, k \in V_{i}}w_{jk}}{\sum_{j\in V_{i}, k \in V}w_{jk}}.
\end{align}
NC is supposed to be minimized since partitions with small NC values have high correlation within groups and low correlation between groups.
According to Eq. (\ref{Eq:NC}) and (\ref{Eq:NA}), $NC(\Gamma_{V}^{K}) + NA(\Gamma_{V}^{K}) = 1$.
Hence, the minimization of NC is equivalent to the maximization of NA.
Without losing generality, we focus on the maximization of NA in the following discussion.

Though the optimization of NC (or NA) has been demonstrated NP-hard \cite{DBLP:journals/pami/ShiM00},
it is much easier to be solved when CBIC is imposed.

According to CBIC,
a graph partition $\Gamma_{V}^{K} = \{V_{1}, V_{2},..., V_{K}\}$
is dependent on a CBIV $\mathbf{s} = (s_{1}, ..., s_{K-1})^{T}$ and there is $V_{i} = [s_{i-1}+1, s_i]$.
In this sense, NA can be rewritten as follows:
\begin{align}
\label{Eq:assoc}
NA(\mathbf{s}) &= \frac{1}{K}\sum_{i=1}^{K}\frac{\sum_{j = s_{i-1}+1}^{s_{i}}\sum_{k = s_{i-1}+1}^{s_{i}} w_{jk}}{\sum_{j = s_{i-1}+1}^{s_{i}} \sum_{k = 1}^{L} w_{jk}}.
\end{align}
As we can see, Eq. (\ref{Eq:assoc}) is a special case of Eq. (\ref{Eq:objective_function}) where the mapping $f$ is defined as:
\begin{equation}
\label{Eq:NC_f}
f_{na}(X_i^j) = \frac{1}{K} \frac{\sum_{k = i}^{j}\sum_{l = i}^{j} w_{kl}}{\sum_{k = i}^{j} \sum_{l = 1}^{L} w_{kl}}.
\end{equation}
Consequently, NA can be finally maximized by inputting $f_{na}$ into Algorithm 1.

One remaining problem is about how to measure the similarities among bands.
Here we adopt a non-parameter method called \emph{local scaling} \cite{DBLP:conf/nips/Zelnik-ManorP04} to construct such a similarity matrix:
\begin{equation}
\label{Eq:local_scaling}
w_{ij} = exp(-\frac{\lVert \mathbf{x}_{i} - \mathbf{x}_{j} \rVert^{2}}{\sigma_{i}\sigma_{j}}),
\end{equation}
where $\sigma_{i} = \lVert \mathbf{x}_{i} - \mathbf{x}_{m} \rVert^{2}$ is the local scaling parameter,
and $\mathbf{x}_{m}$ is the $m$-th neighbor of $\mathbf{x}_{i}$ ($m$ is set to $7$ according to \cite{DBLP:conf/nips/Zelnik-ManorP04}).

\textbf{Top-rank cut criterion}.
Except for NC,
a novel \emph{top-rank cut} (TRC) criterion is proposed in this paper.
There are two motivations to design TRC.
1) Sometimes, a band interval with minimal contribution to the value of objective function has even negative effect on an algorithm's performance.
This stems from the fact that there may exist noisy bands in that interval.
Hence, to maximize this minimal contribution among all the band intervals may be a more effective way to attain the clustering result.
2) Since in the proposed RCS,
only the bands with the highest rank values will be selected in each subset.
Those bands are more important and should have larger priorities when designing the objective function.

According to Section \ref{sec:E-OCF},
we will give the definition of TRC by giving the tuple $(\oplus, f)$,
where the binary operator $\oplus$ is the maximization operation: $\forall a,b \in \mathbb{R}, a \oplus b = \max(a,b)$,
and the mapping $f$ is defined as:
\begin{equation}
\label{Eq:TRC_f}
   f_{trc}(X_i^j) = \sum_{k \in [1,i) \cup (j,L]}w_{pk},
   \quad \mathbf{s.t.}\quad p = \mathop{\arg\max}\limits_{i \leq l \leq j} r_l,
\end{equation}
in which $r_l$ is the rank value of $\mathbf{x}_l$ calculated when conducting RCS.

To explain with it, TRC characterizes a band partition in the following steps:
1) Choose the bands with the highest rank values in each cluster.
2) For each chosen band $\mathbf{x}_p$,
calculate the sum of similarities between $\mathbf{x}_p$ and the other bands out of the cluster it belongs to.
This sum value is the score of the corresponding cluster.
3) The maximum score among all the clusters is just the TRC value of this band partition.
It should be pointed out that, different from NC, the value of TRC should be minimized to achieve a promising result.
TRC is easy to be optimized through some minor modifications of Algorithm \ref{Alg:OCF}.

\subsection{Ranking Methods for RCS}
\label{sec:Implementation:RCS}
To conduct the proposed RCS, a ranking method is needed to evaluate the importance of bands.
In this subsection, some ranking criteria will be presented and discussed to accomplish this task.

1) MVPCA \cite{DBLP:journals/tgrs/ChangDSA99}.
As stated in Section \ref{sec:related_work}, MVPCA evaluate the bands according to their variances.
Generally, low variance means that different ground objects have similar characteristic,
and are difficult to be recognized.
In contrary, large variance represents for more distinctive information,
so that ground objects are easier to be distinguished.
Unfortunately, one disadvantage of MVPCA is that it is sensitive to noises
since noisy bands usually have large variances.

2) E-FDPC \cite{7161371}. E-FDPC ranks each band by assessing whether it is a suitable cluster center.
First, a cluster center should have large local density, i.e.,
there should be lots of bands close to it.
Second, it should be distant from the bands that have larger local densities than it.
For E-FDPC, the first property eliminates the noisy bands since they usually have low local density,
while the second property helps to reduce the correlation among bands.

3) Information entropy (IE) \cite{Shannon1948A}.
IE is a noise insensitive criterion to measure the information hidden in a stochastic variable.
For a band $\mathbf{x}_l$, its entropy can be defined as:
\begin{equation}
\label{Eq:TRC}
H(\mathbf{x}_l) = -\sum_{\omega\in \Omega}p(\omega)\ {\rm log}\ p(\omega),
\end{equation}
where $\Omega$ is the gray-scale color space,
and $p(\omega)$ can be calculated according to the gray-level histogram of $\mathbf{x}_l$ \cite{7888954, 7214263}.

Until now, we have listed all the objective functions and ranking methods utilized in this paper.
For convenient expression, the devised algorithm will be named following the pattern ``objective function''-OC-``ranking method'', e.g.,
TRC-OC-FDPC denotes that TRC and E-FDPC are utilized to implement the algorithm.

\subsection{Analysis of Computational Complexity}
The conduction of the proposed algorithm can be decomposed into $3$ steps,
they are pre-processing step, clustering step and selection step.
In the following, we will discuss the theoretical computational complexity for each step.

\textbf{Pre-processing step}.
In this step, we seek to calculate the mapping $f$ for each kind of objective function.
For both NC and TRC, a similarity matrix should be first calculated according to Eq. (\ref{Eq:local_scaling}),
and this takes $O(L^2N)$ computational complexity.
For TRC, each band should be further aligned with rank values,
whose complexity is depending on the utilized ranking methods (which will be discussed in the third step).
Based on these preparatory works, the mapping $f$ can be calculated according to Eq. (\ref{Eq:NC_f}) and ($\ref{Eq:TRC_f}$)
by taking $O(L^3)$ complexity for both NC and TRC.

\textbf{Clustering step}.
In this step, the optimal CBIV $\mathbf{s}^*$ is obtained by optimizing the objective functions.
According to Algorithm 1, $O(L^2K)$ time is needed for both NC and TRC.

\textbf{Selection step}.
In the last step, bands should be assigned with rank values to conduct RCS.
For MVPCA, E-FDPC and IE, their computational complexities are $O(LN)$, $O(L^2N)$ and $O(LN)$, respectively.

Since $K < L \ll N$,
we know that all versions of the proposed algorithm cost $O(L^2N)$ time.
This is acceptable for most of the applications.
One should be noted that,
the efficiency of the proposed framework is limited by the pre-processing step and selection step.
The clustering step, which is considered as the main step, is rather more efficient.
Hence we believe the proposed framework is potential to be suitable for
more time-critical applications only if simpler objective functions and ranking methods are utilized.

\subsection{Determine the Number of the Selected Band}
\label{ssec:determin_number}
In \cite{DBLP:journals/tgrs/ChangDSA99}, a variance-based band-power ratio is defined to determine the required number of bands.
It first calculates the variances $\sigma^2_l$ for each $1\leq l \leq L$, and sort them in descending order.
Then define a band-power ratio:
\begin{equation}
    R_{var}(k) = \frac{\sum_{l=1}^{k}\sigma^2_l}{E},
\end{equation}
where $E=\sum_{l=1}^L\sigma^2_l$ is the sum of variances of all the bands.
According to this ratio,
we can know how many bands is needed to produce a certain percentage of band-power.
However in HSI, bands with large variances may still be strongly correlated,
so there is common part among band power they produce.
With consideration of this characteristic,
a improved method is proposed to measure how much power can be produced by $k$ low-correlated bands.

We first estimate a upper bound of the required number of band $M$,
which is given by $M=\lambda  L$, where $\lambda$ is a ratio between $0$ and $1$.
Suppose $i_1, i_2,..., i_M$ is the indexes of $M$ selected bands by NC-OC-MVPCA (the proposed band selection algorithm using NC and MVPCA).
Without loss of generality, we assume there is $\sigma_{i_1} > \sigma_{i_2}>...>\sigma_{i_M}$.
Then, the correlation-reduced band-power ratio is formulized as:
\begin{equation}
\label{Eq:determine_number}
    R_{crvar}(k) = \frac{\sum_{l=1}^{k}\sigma^2_{i_l}}{E_{cr}},
\end{equation}
where $E_{cr} = \sum_{l=1}^{M}\sigma^2_{i_l}$.
Since the correlation among bands has been reduced in clustering process,
this method can better characterize the distinct information hidden in different bands.
Finally, given a desired correlation-reduced band-power ratio $R^*$,
we can determine the required number of bands $K^*$ by $R_{crvar}(K^*-1) \leq R^* < R_{crvar}(K^*)$.

\section{EXPERIMENT}
\label{sec:experiment}
In this section, comparative experiments are conducted on different HSI data sets.
First, we introduce the experimental setups, including the employed data sets,
comparison methods, classification settings and the number of the required bands.
Then the classification results of all the four data sets are shown and analysed.
Finally the computational times of different methods are compared.

\subsection{Experimental Setup}

\textbf{Data set}.
Four real-world HSI data sets captured by two different image systems are used in the experiments.
They are introduced as follows.

1) Indian Pines Scene. Indian Pines Scene was captured by AVIRIS sensor in North-Western Indiana in 1992.
It consists of $145 \times 145$ pixels and $220$ spectral reflectance bands in the wavelength range of $0.4-2.5 \mu$m.
There are 16 classes of objects contained in the image.
Water absorbtion bands including $104-108$, $150-163$ and $220$ are removed
and a total of $200$ bands are utilized.

2) Pavia University Scene. Pavia University Scene was acquired by the Reflective Optics System Imaging Spectrometer (ROSIS) system during a flight campaign over Pavia,
Northern Italy in 2002.
After some pixels with no information discarded, an image of size $610 \times 340$ is used.
Pavia University Scene has $103$ bands and $9$ classes of land cover objects.

3) Salinas Scene. Salinas Scene was captured by AVIRIS sensor in Salinas Valley, California in 1998.
The image size of Salinas Scene is $512 \times 217$ with a spectral coverage within $0.4-2.5\mu$m.
There also $224$ spectral bands and $16$ classes of interests in the image.
Similarly, $20$ water absorption bands are discarded including $108-112$, $154-167$ and $224$,
and finally a total of $204$ bands are used in the experiments.

4) Kennedy Space Center (KSC). KSC was captured by AVIRIS sensor in Florida, on March 23, 1996.
It was acquired from an altitude of approximately $20$km, with a spatial resolution of $18$m.
there are total $176$ bands after removing $48$ bands with low SNR or water absorption,
and each band is with size $512 \times 614$ and has $13$ classes of land cover objects.

\textbf{Comparison method}.
To verify the effectiveness of the proposed algorithms,
several state-of-the-art methods are included as competitors.
They are WaLuDi \cite{DBLP:journals/tgrs/UsoPSG07},
uniform band selection (UBS) \cite{DBLP:journals/tgrs/ChangW06},
volume gradient band selection (VGBS) \cite{DBLP:journals/tgrs/GengSJZ14},
enhanced fast density-peak-based clustering (E-FDPC) \cite{7161371}, and
multi-task sparsity pursuit (MTSP) \cite{DBLP:journals/tgrs/YuanZW15}.
About the parameters among them,
WaLuDi, UBS, VGBS and E-FDPC are parameter-free so only $K$ is needed to be set.
For MTSP, its parameters are tuned on Indian Pines and fixed for the other three data sets.
Note that the proposed framework itself are parameter-free.
The only parameters that needed to be set are $K$ and the number of the selected bands involved by E-FDPC, denoted as $K^{'}$.
Considering that E-FDPC is used to prioritize all the $L$ bands, $K^{'}$ is fixed to $L$ for all the data sets.


\textbf{Classification setting}.
Four classifiers are utilized to examine the classification accuracies of different band selection methods.
They are k-nearest neighborhood (KNN) \cite{DBLP:journals/tit/CoverH67},
linear discriminant analysis (LDA) \cite{DBLP:journals/tgrs/BandosBC09},
support vector machine (SVM) \cite{DBLP:journals/tgrs/MelganiB04},
and edge-preserving filtering (EPF) \cite{6553593}.
In the experiments, $10\%$ of the samples for each class are chosen randomly to train the classifiers, while the rest are used in testing.
To simplify the problem, background is not considered.
In order to reduce the instability caused by the random selection of the training samples,
the final results are achieved by averaging $10$ individual runs.

\begin{figure}[htb]
\begin{minipage}[b]{1.0\linewidth}
  \centering
  \centerline{\epsfig{figure=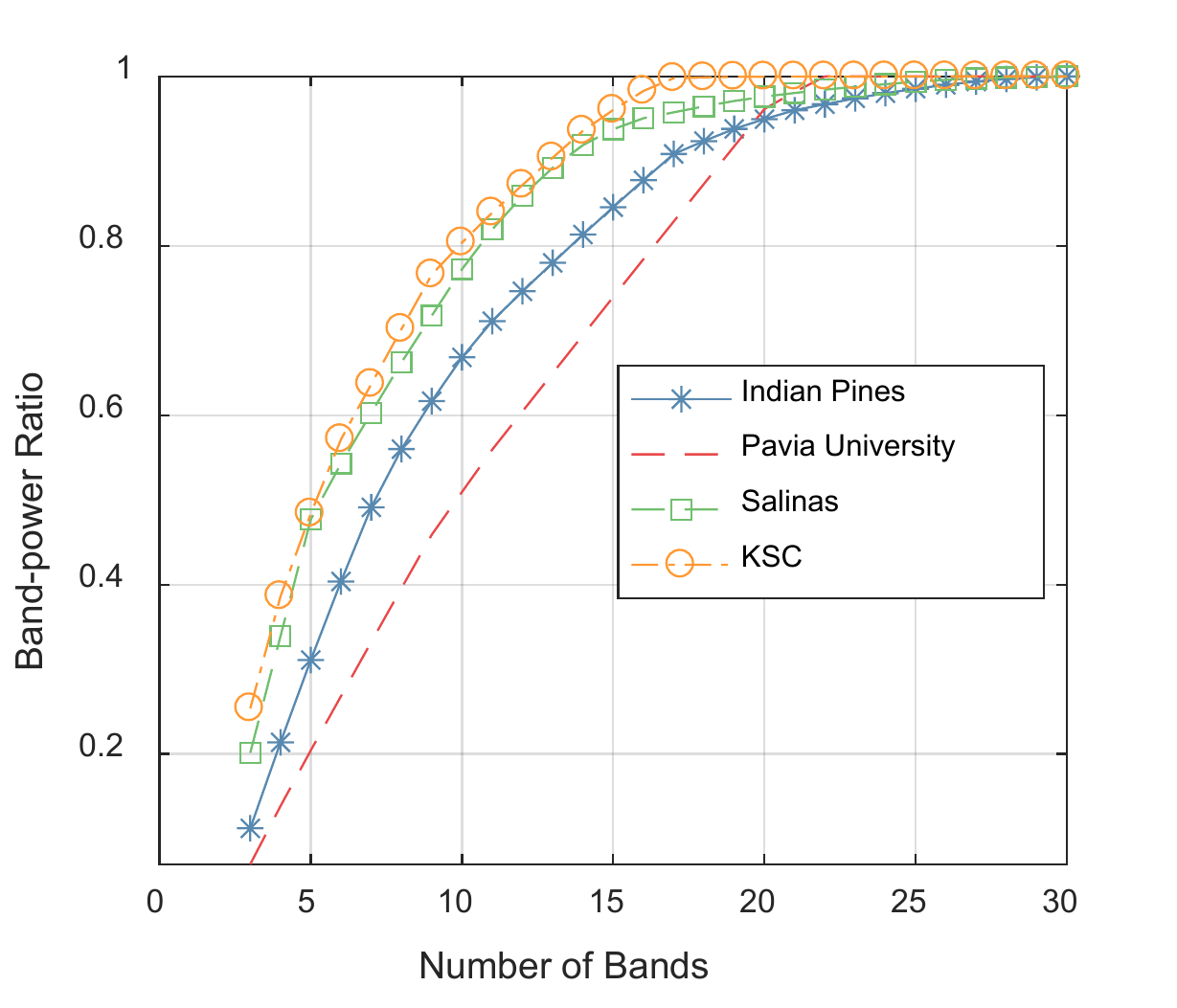,width=7cm}}
\end{minipage}
\caption{The values of $R_{crvar}$ against the number of bands on different data sets.}
\label{fig:determine_number}
\end{figure}

\textbf{Number of the required bands}.
To determine the number of the required bands,
the upper bound $M$ in Section \ref{ssec:determin_number} is set to $\frac{L}{5}$,
and $R^*$ is set to $0.8$ empirically for all the four data sets.
Function $R_{crvar}$ are plotted in Fig. \ref{fig:determine_number}.
According to Eq. (\ref{Eq:determine_number}),
the numbers of the required bands are $14$, $17$, $11$ and $10$ for Indian Pines, Pavia University, Salinas and Kennedy Space Center, respectively.

\subsection{Result Analysis}
\label{ssec:classification_result}
In this part, we compare the proposed algorithms with some state-of-the-art methods.
Overall accuracy (OA) is involved as the evaluation criterion.
In each data set, three indicators will be compared.
They are:
1) OA curves, i.e., OA values versus the number of the selected band $K$ ($K$ is set every $3$ intervals from $3$ - $30$).
2) Average OA against different number of bands.
3) OA value when the number of bands is determined by Eq. (\ref{Eq:determine_number}).
In the experiments,
four versions of the proposed algorithm will be examined.
They are TRC-OC-FDPC, NC-OC-IE, TRC-OC-IE and NC-OC-MVPCA.
For simplicity, only the first two of them will be shown in OA curves.
For comparison purpose, $14$ selected bands for all the methods in Indian Pines are listed in Table \ref{table:selected_bands}
(their indexes are that before the removal of noisy bands).

\begin{figure}[htb]
\begin{minipage}[b]{1.0\linewidth}
  \centering
  \centerline{\epsfig{figure=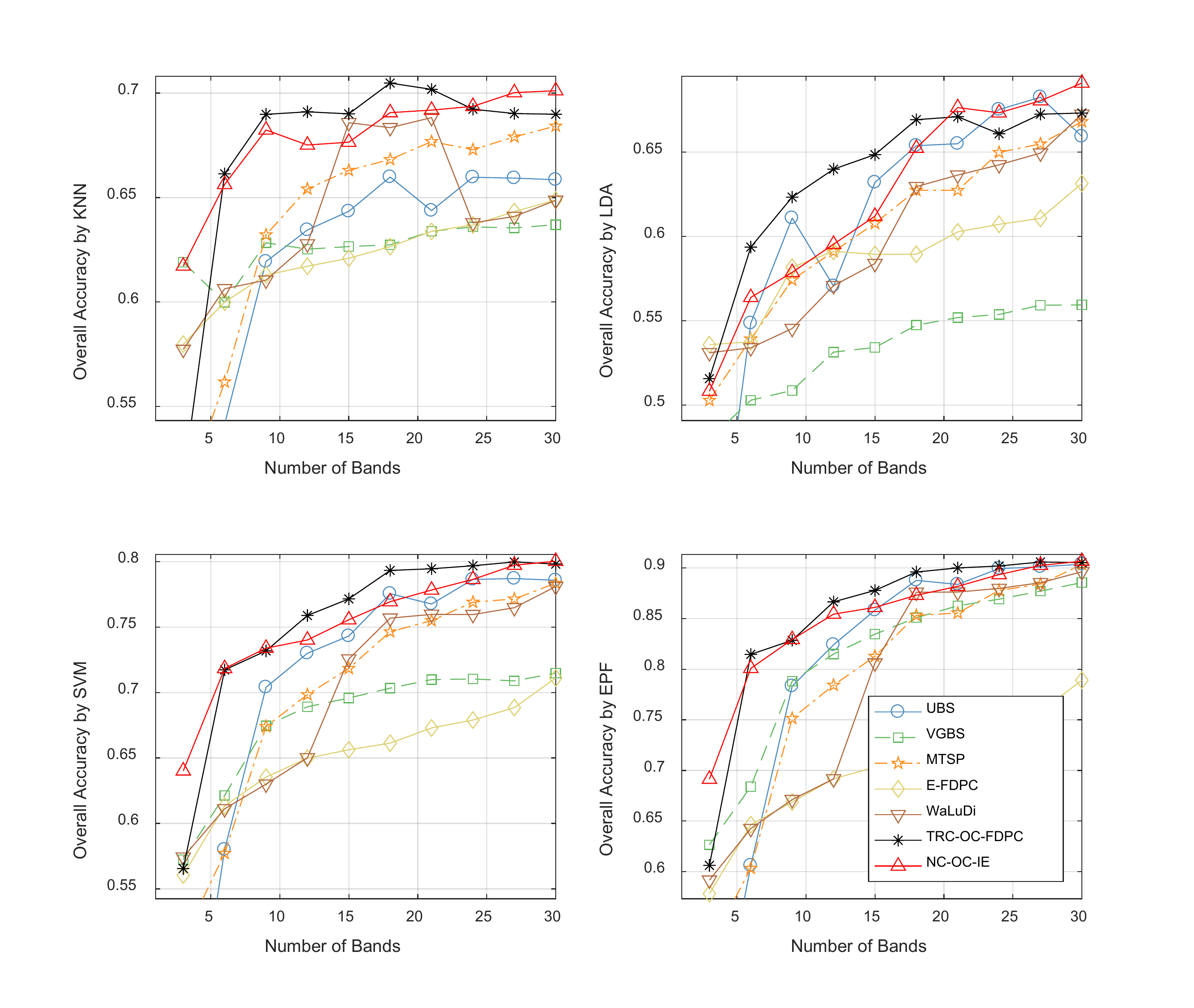,width=8.5cm}}
\end{minipage}
\caption{OA curves on Indian Pines Scene for different band selection methods.}
\label{fig:Indian_Pines_OACurves}
\end{figure}

\begin{figure}[htb]
\begin{minipage}[b]{1.0\linewidth}
  \centering
  \centerline{\epsfig{figure=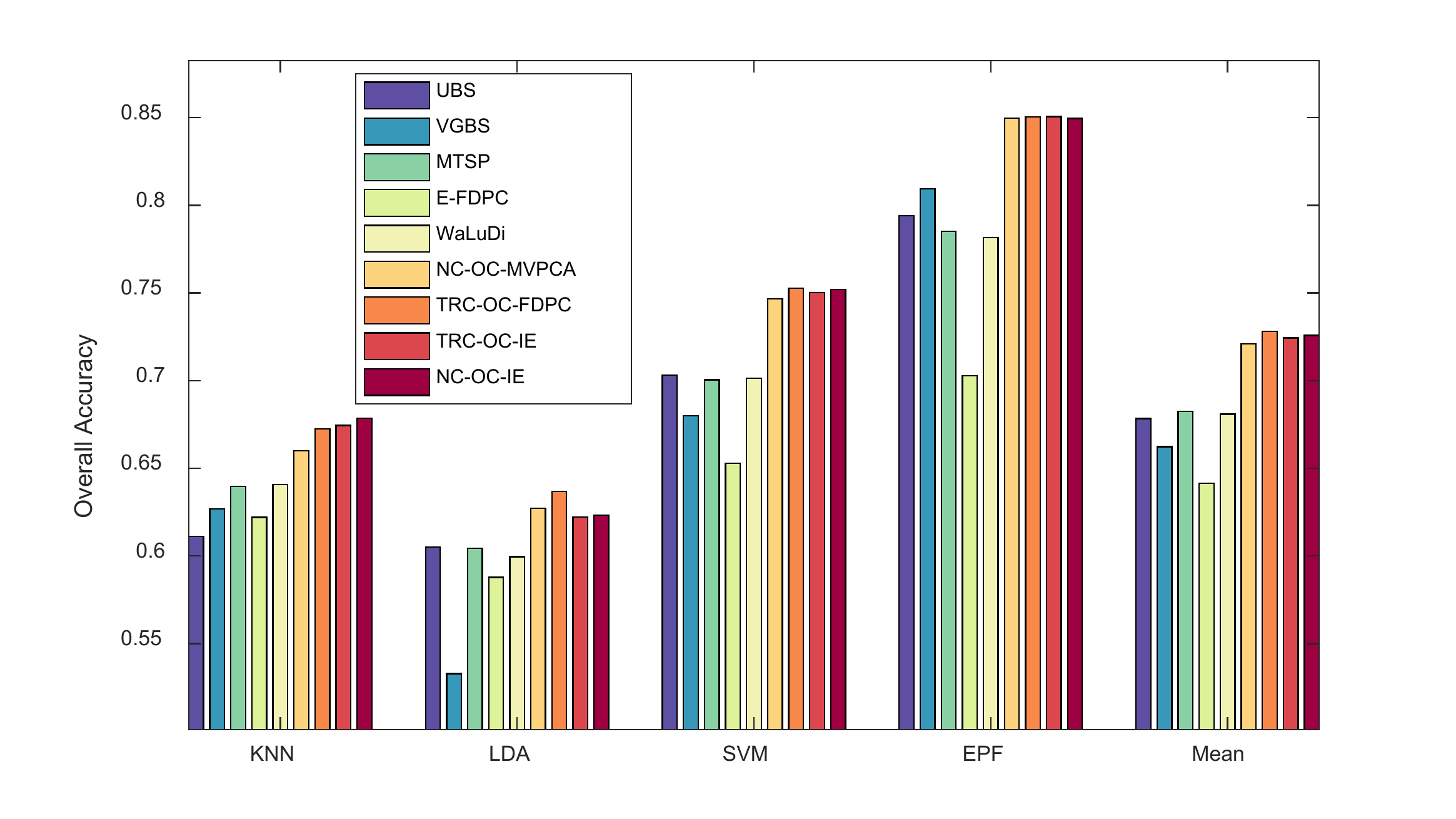,width=8.0cm}}
\end{minipage}
\caption{Average OA values against the first $30$ bands on Indian Pines Scene for different band selection methods.}
\label{fig:Indian_Pines_MeanOA}
\end{figure}

\begin{figure}[htb]
\begin{minipage}[b]{1.0\linewidth}
  \centering
  \centerline{\epsfig{figure=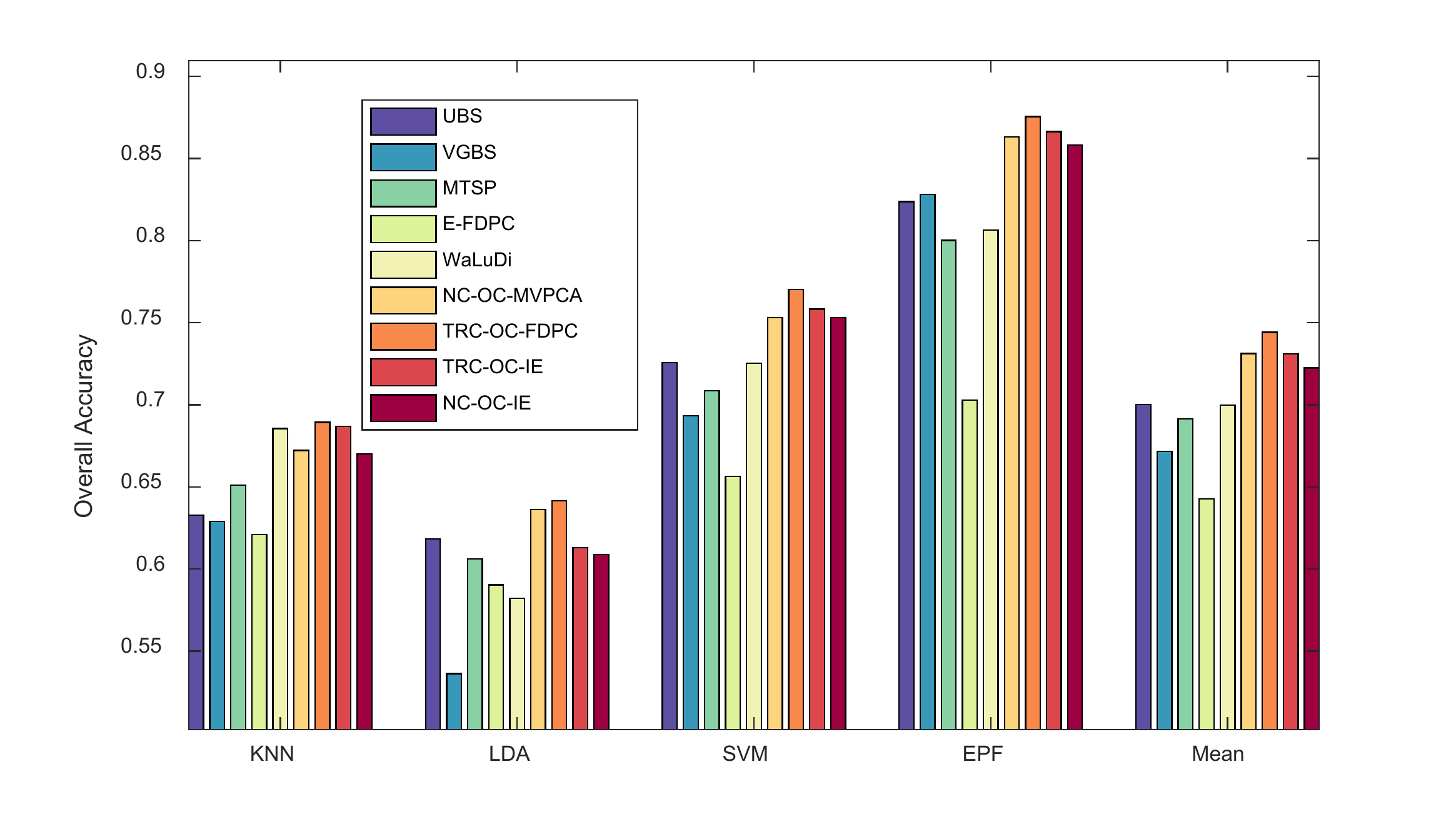,width=8.0cm}}
\end{minipage}
\caption{OA values on Indian Pines Scene for different band selection methods when $14$ bands are selected.}
\label{fig:Indian_Pines_OANumber}
\end{figure}

\textbf{Indian Pines Scene}.
Fig. \ref{fig:Indian_Pines_OACurves}, \ref{fig:Indian_Pines_MeanOA} and \ref{fig:Indian_Pines_OANumber} show
the above three kinds of indicators on Indian Pines using different classifiers.
In Fig. \ref{fig:Indian_Pines_OACurves},
we can see TRC-OC-FDPC and NC-OC-IE achieve the best performance,
with stable and high OA values against different number of bands and classifiers.
When refer to Fig. \ref{fig:Indian_Pines_MeanOA},
the proposed algorithms achieve higher average OAs compared to the others.
Fig. \ref{fig:Indian_Pines_OANumber} shows the OA values when $14$ bands are selected.
The proposed algorithms also show significant superiority in this case.


\begin{table}[htbp] \small
    \centering
        \caption{$14$ selected bands on Indian Pines data set.}
            \begin{tabular}{cc}
                \toprule
                Method Names       & $14$ Selected Bands\\ \midrule
                TRC-OC-FDPC & 8/16/28/30/43/50/67/92/123/133/\\&146/176/182/188/       \\
                NC-OC-IE    & 17/29/42/48/54/119/122/137/146/179/\\&185/195/204/213/ \\
                TRC-OC-IE   & 15/21/29/42/54/79/89/116/122/127/\\&146/147/179/201/     \\
                NC-OC-MVPCA & 17/29/42/48/57/119/122/137/173/180/\\&185/195/204/213/  \\
                UBS          & 1/16/31/46/62/77/92/113/128/143/\\&173/188/203/219/    \\
                VGBS         & 1/13/18/20/23/29/34/35/39/57/\\&61/75/88/89/          \\
                MTSP         & 12/13/29/37/44/45/81/90/112/141/\\&179/186/202/215/     \\
                E-FDPC       & 50/67/80/92/105/109/118/124/134/137/\\&142/152/167/179/\\
                WaLuDi       & 31/49/56/61/67/77/83/99/104/128/\\&138/163/183/189/    \\
                \bottomrule
            \end{tabular}
\label{table:selected_bands}
\end{table}

\begin{figure}[htb]
\begin{minipage}[b]{1.0\linewidth}
  \centering
  \centerline{\epsfig{figure=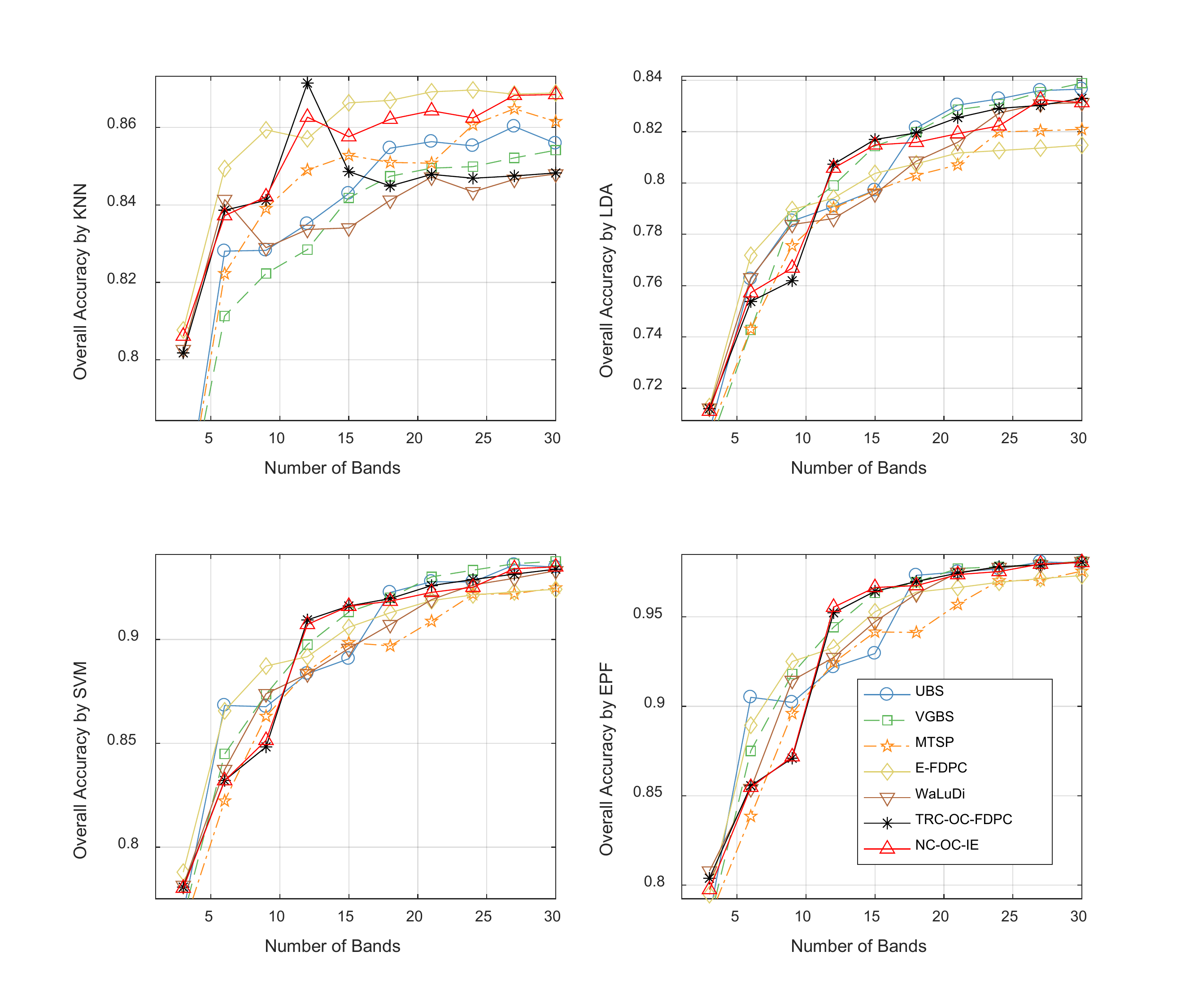,width=8.5cm}}
\end{minipage}
\caption{OA curves on Pavia University Scene for different band selection methods.}
\label{fig:Pavia_University_OACurves}
\end{figure}

\begin{figure}[htb]
\begin{minipage}[b]{1.0\linewidth}
  \centering
  \centerline{\epsfig{figure=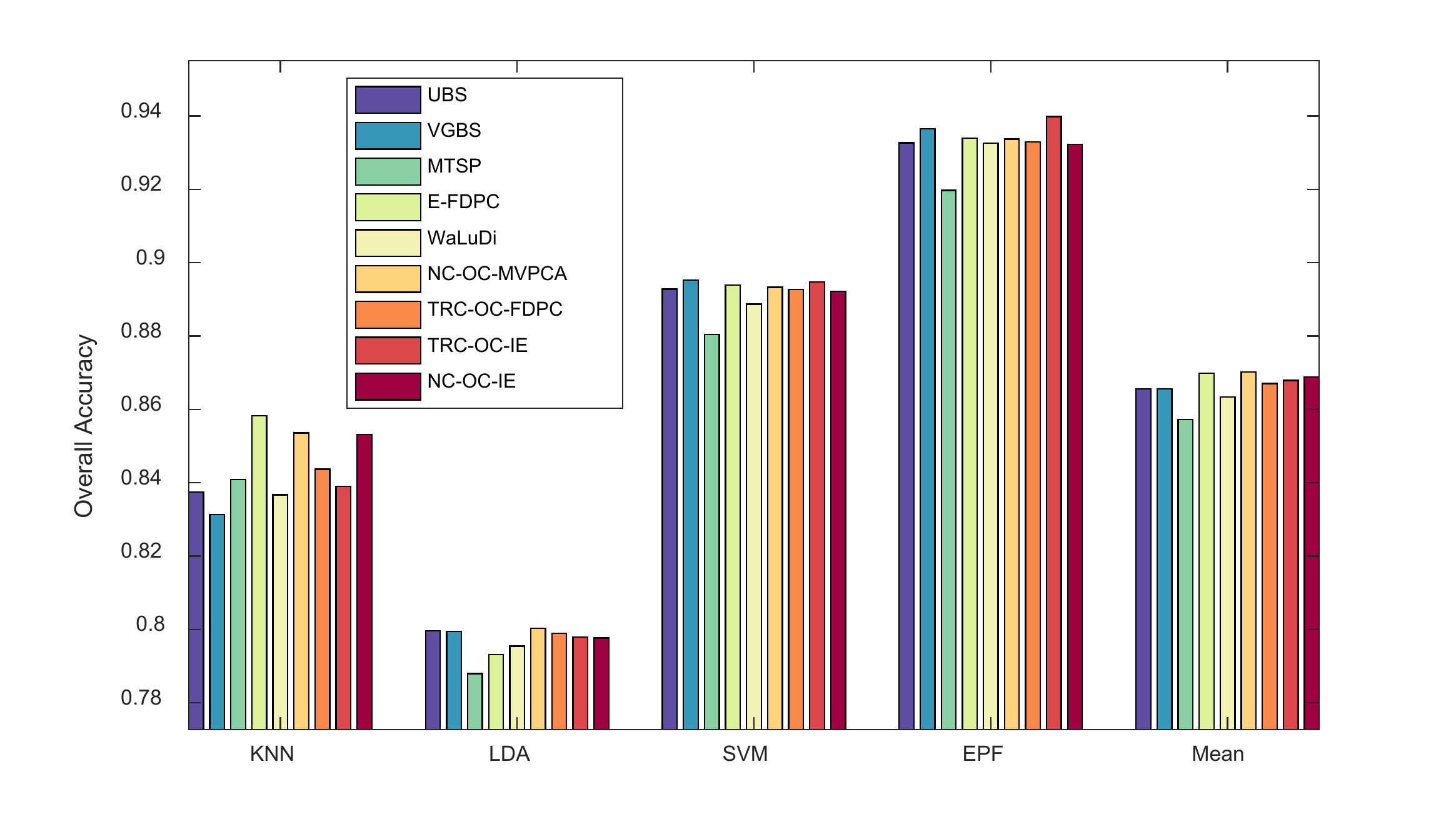,width=8.0cm}}
\end{minipage}
\caption{Average OA values against the first $30$ bands on Pavia University Scene for different band selection methods.}
\label{fig:Pavia_University_MeanOA}
\end{figure}

\begin{figure}[htb]
\begin{minipage}[b]{1.0\linewidth}
  \centering
  \centerline{\epsfig{figure=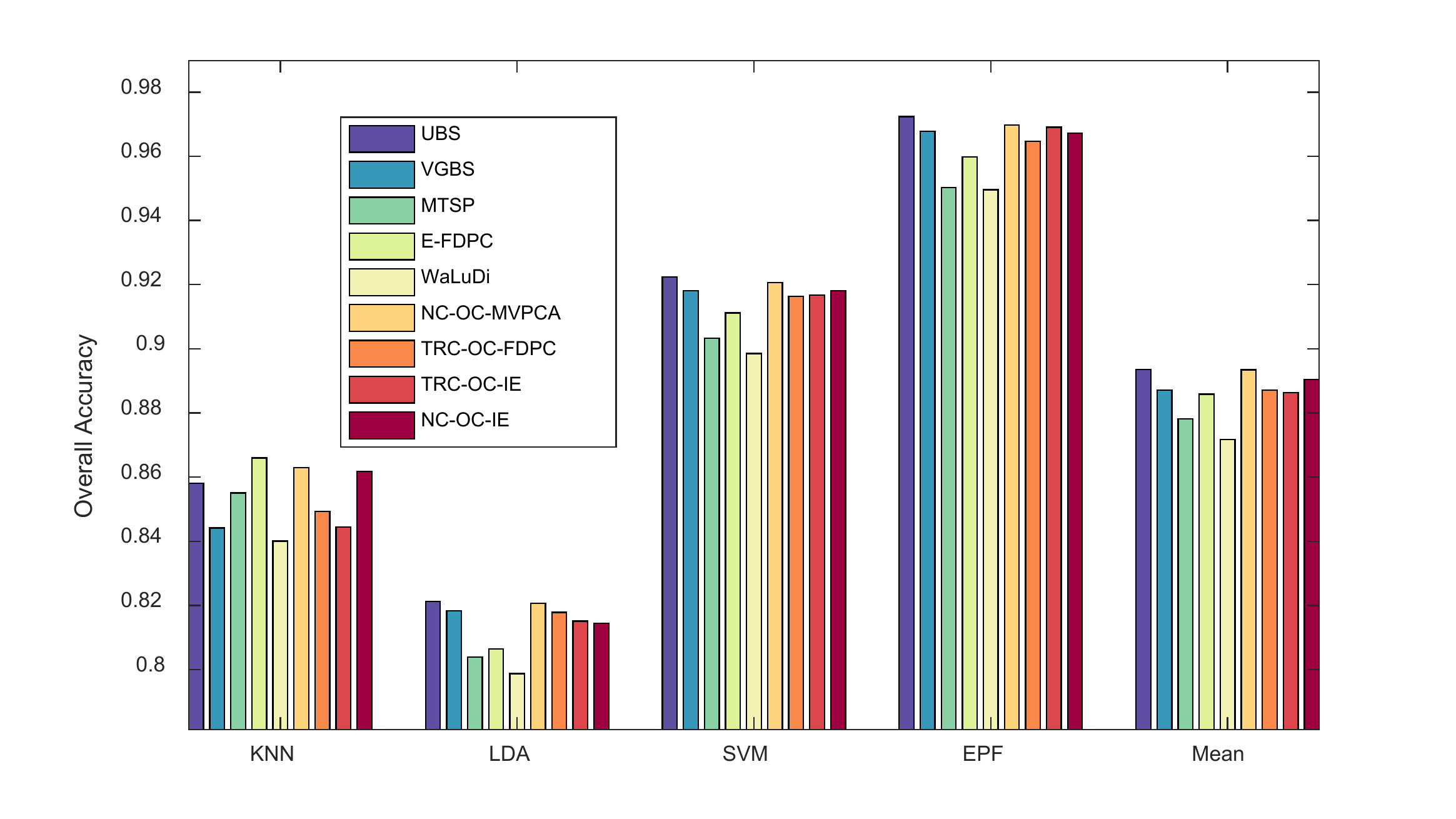,width=8.0cm}}
\end{minipage}
\caption{OA values on Pavia University Scene for different band selection methods when $17$ bands are selected.}
\label{fig:Pavia_University_OANumber}
\end{figure}

\textbf{Pavia University Scene}.
Similar to Indian Pines Scene, the above three indicators are shown in
Fig. \ref{fig:Pavia_University_OACurves}, \ref{fig:Pavia_University_MeanOA} and \ref{fig:Pavia_University_OANumber}.
In Fig. \ref{fig:Pavia_University_OACurves}, the results are different with respect to classifiers.
While KNN is employed, NC-OC-IE attains a stable OA curve but
TRC-OC-FDPC acquires a large decrease when $15$ bands are selected.
For other classifiers, they both perform poor when the number of bands is $6$ or $9$,
but are superior to all the competitors when it is $12$ or $15$.
In Fig. \ref{fig:Pavia_University_MeanOA},
the difference among methods is very small.
When averaging the results produced by all the classifiers, NC-OC-MVPCA slightly outperforms E-FDPC and ranks the first.
In Fig. \ref{fig:Pavia_University_OANumber} when $17$ bands are selected,
UBS and NC-OC-MVPCA outperform the others on LDA, SVM and EPF,
while E-FDPC performs better on KNN.
When averaging the results,
NC-OC-MVPCA and UBS achieve the best performance.

\begin{figure}[htb]
\begin{minipage}[b]{1.0\linewidth}
  \centering
  \centerline{\epsfig{figure=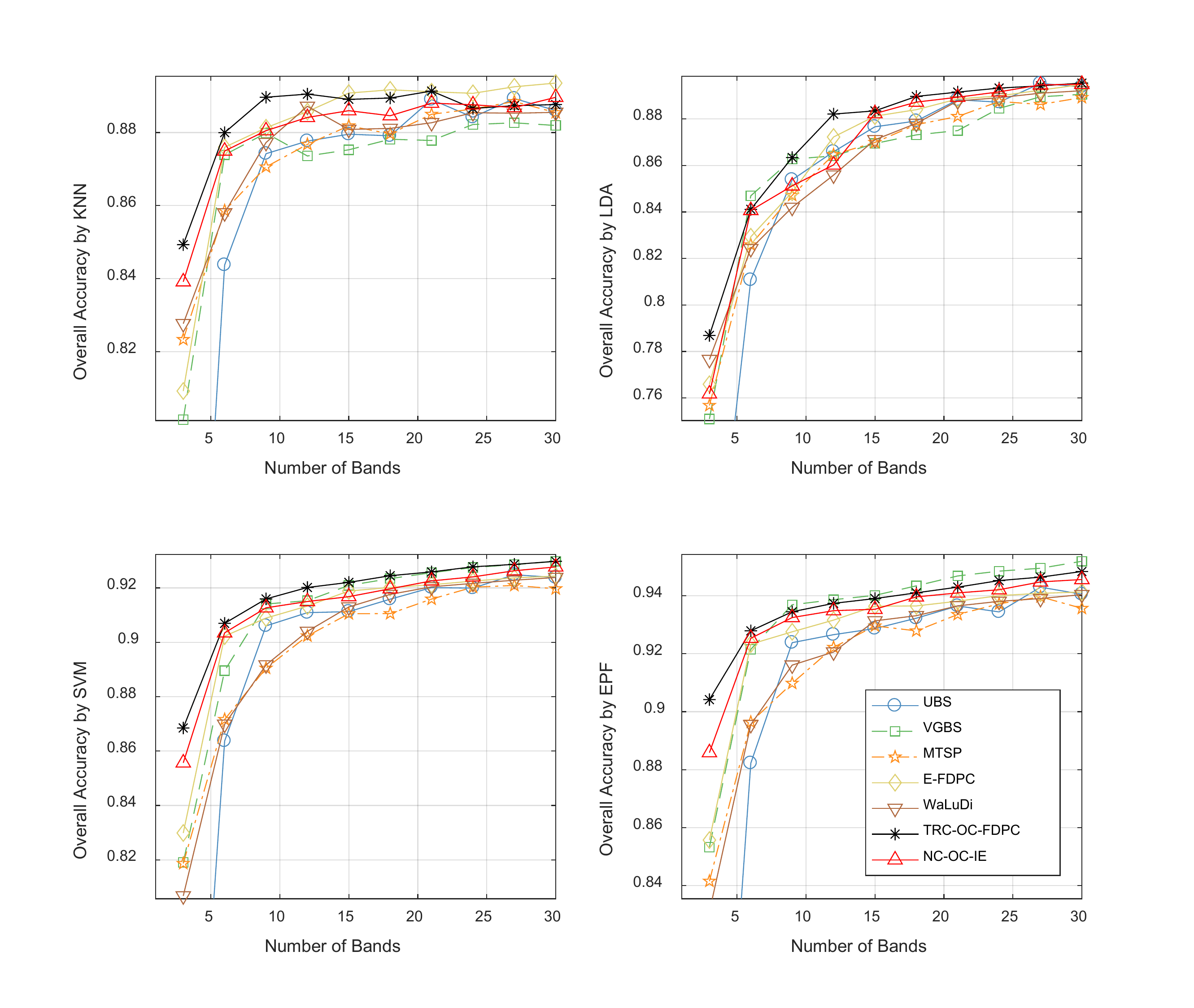,width=8.5cm}}
\end{minipage}
\caption{OA curves on Salinas Scene for different band selection methods.}
\label{fig:Salinas_OACurves}
\end{figure}

\begin{figure}[htb]
\begin{minipage}[b]{1.0\linewidth}
  \centering
  \centerline{\epsfig{figure=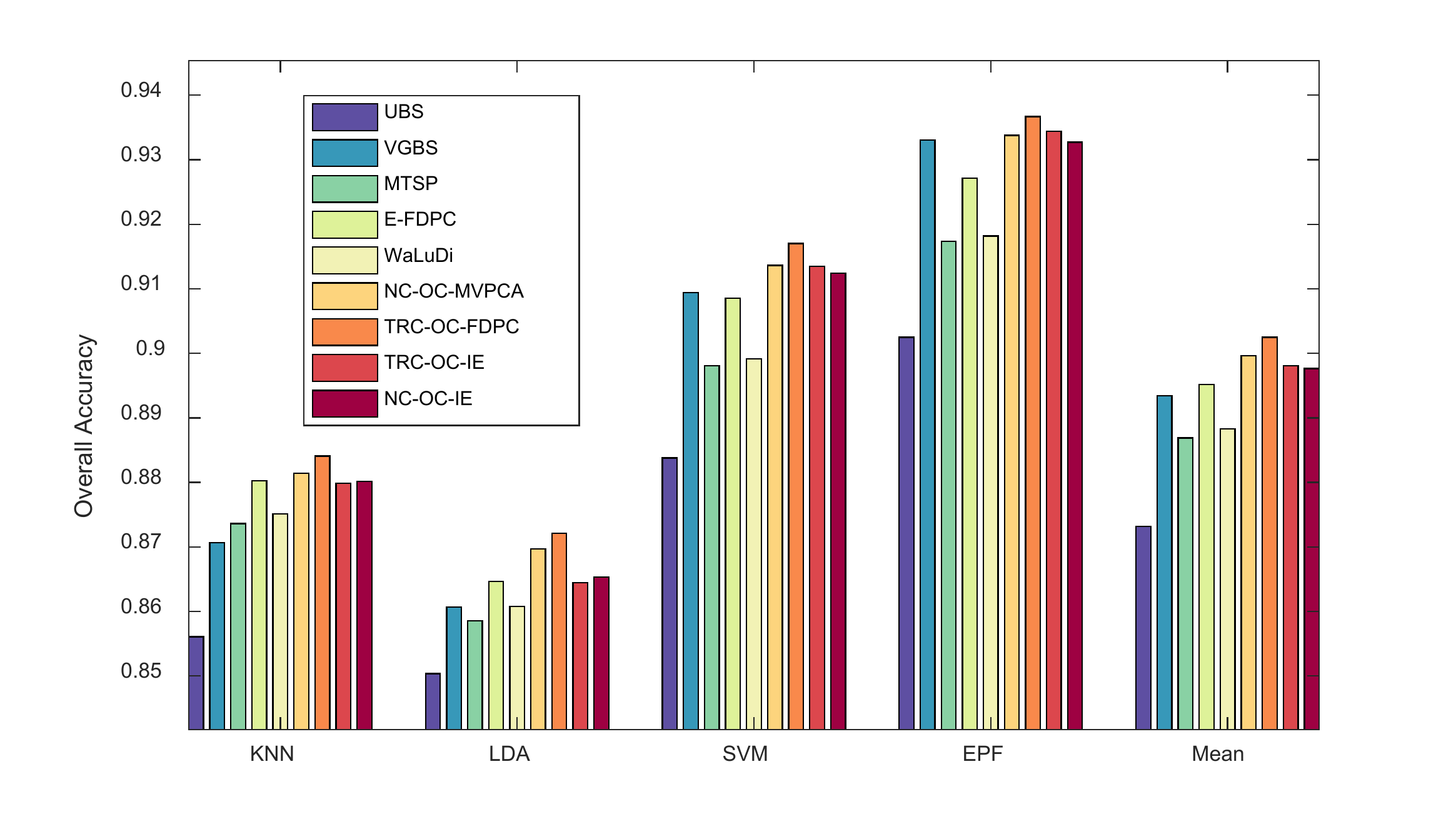,width=8.0cm}}
\end{minipage}
\caption{Average OA values against the first $30$ bands on Salinas Scene for different band selection methods.}
\label{fig:Salinas_MeanOA}
\end{figure}

\begin{figure}[htb]
\begin{minipage}[b]{1.0\linewidth}
  \centering
  \centerline{\epsfig{figure=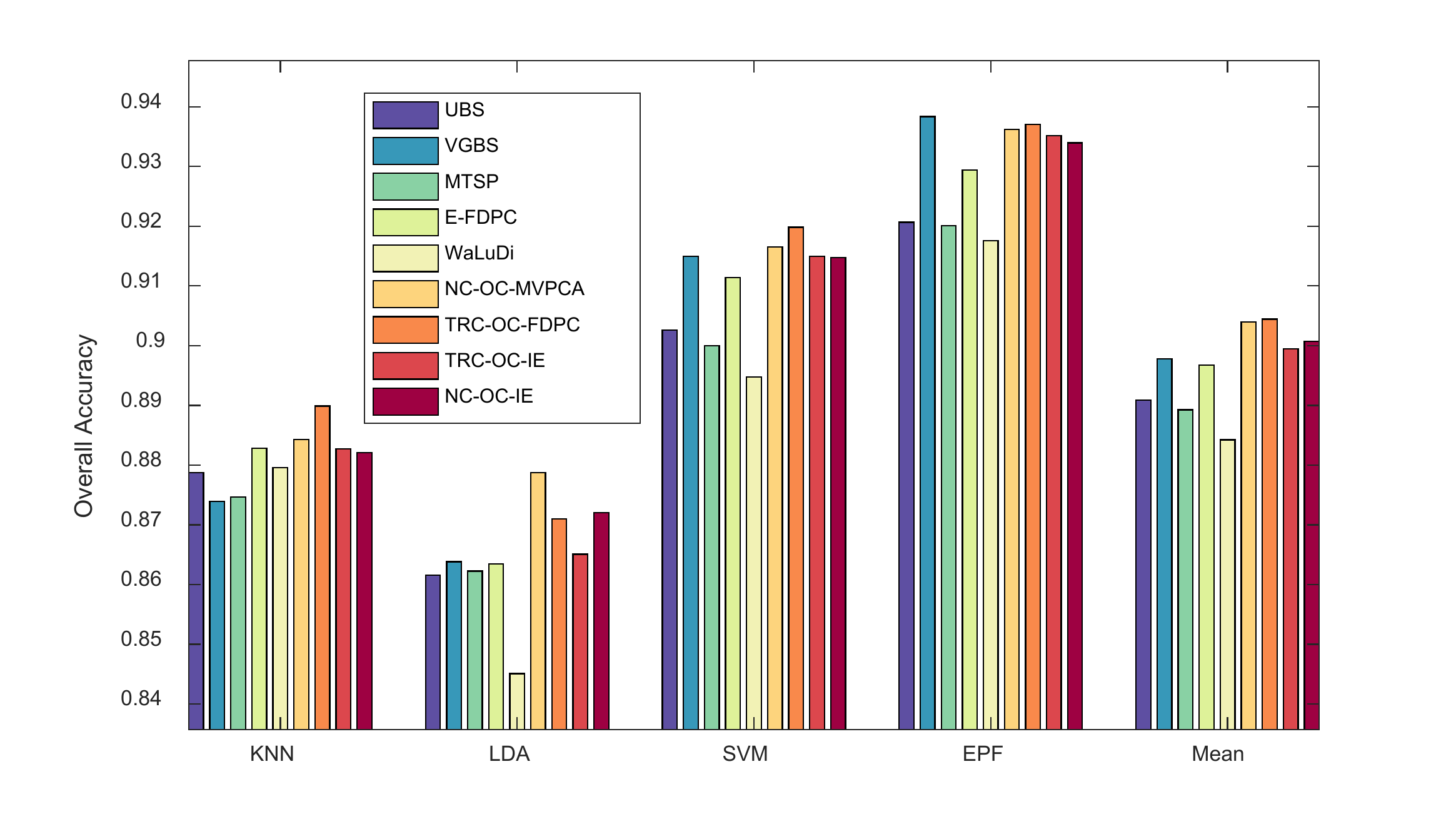,width=8.0cm}}
\end{minipage}
\caption{OA values on Salinas Scene for different band selection methods when $11$ bands are selected.}
\label{fig:Salinas_OANumber}
\end{figure}

\textbf{Salinas Scene}.
For Salinas Scene, the above three indicators are shown in Fig. \ref{fig:Salinas_OACurves}, \ref{fig:Salinas_MeanOA} and \ref{fig:Salinas_OANumber}.
In Fig. \ref{fig:Salinas_OACurves},
TRC-OC-FDPC achieves a satisfactory performance,
and is superior to the other methods in general.
NC-OC-IE also achieves good performance.
It is more robust against various classifiers compared to the other $5$ competitors.
When referring to the average OA value as shown in Fig. \ref{fig:Salinas_MeanOA},
the proposed algorithms are more effective than the others in most of the cases.
In Fig. \ref{fig:Salinas_OANumber},
the proposed algorithms are inferior to VGBS when EPF is employed,
but are still more superior to the others for other classifiers.

\begin{figure}[htb]
\begin{minipage}[b]{1.0\linewidth}
  \centering
  \centerline{\epsfig{figure=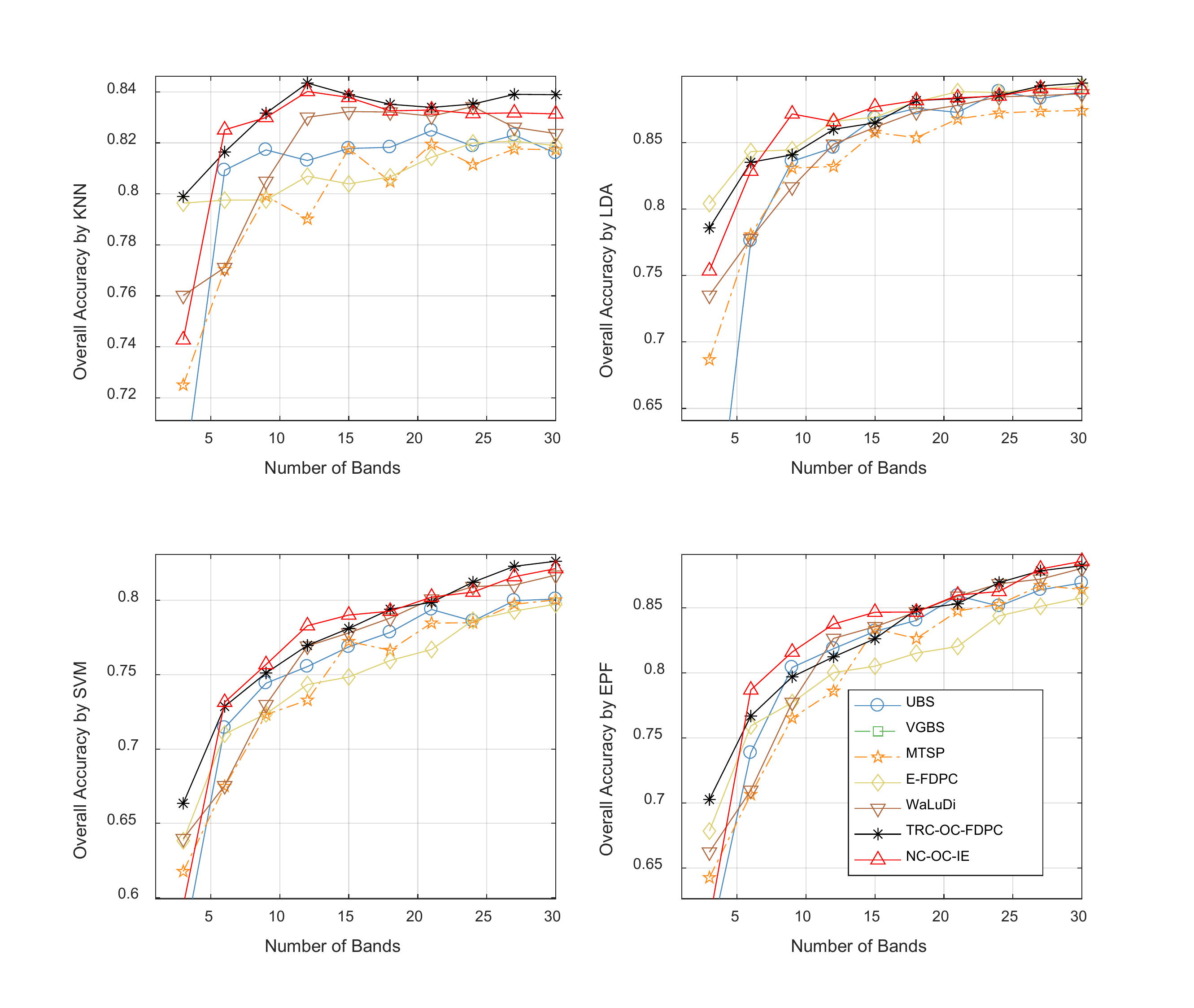,width=8.5cm}}
\end{minipage}
\caption{OA curves on KSC for different band selection methods.
(The OA values of VGBS are lower than the lower bound of y-axes, so they cannot be seen.)}
\label{fig:KSC_OACurves}
\end{figure}

\begin{figure}[htb]
\begin{minipage}[b]{1.0\linewidth}
  \centering
  \centerline{\epsfig{figure=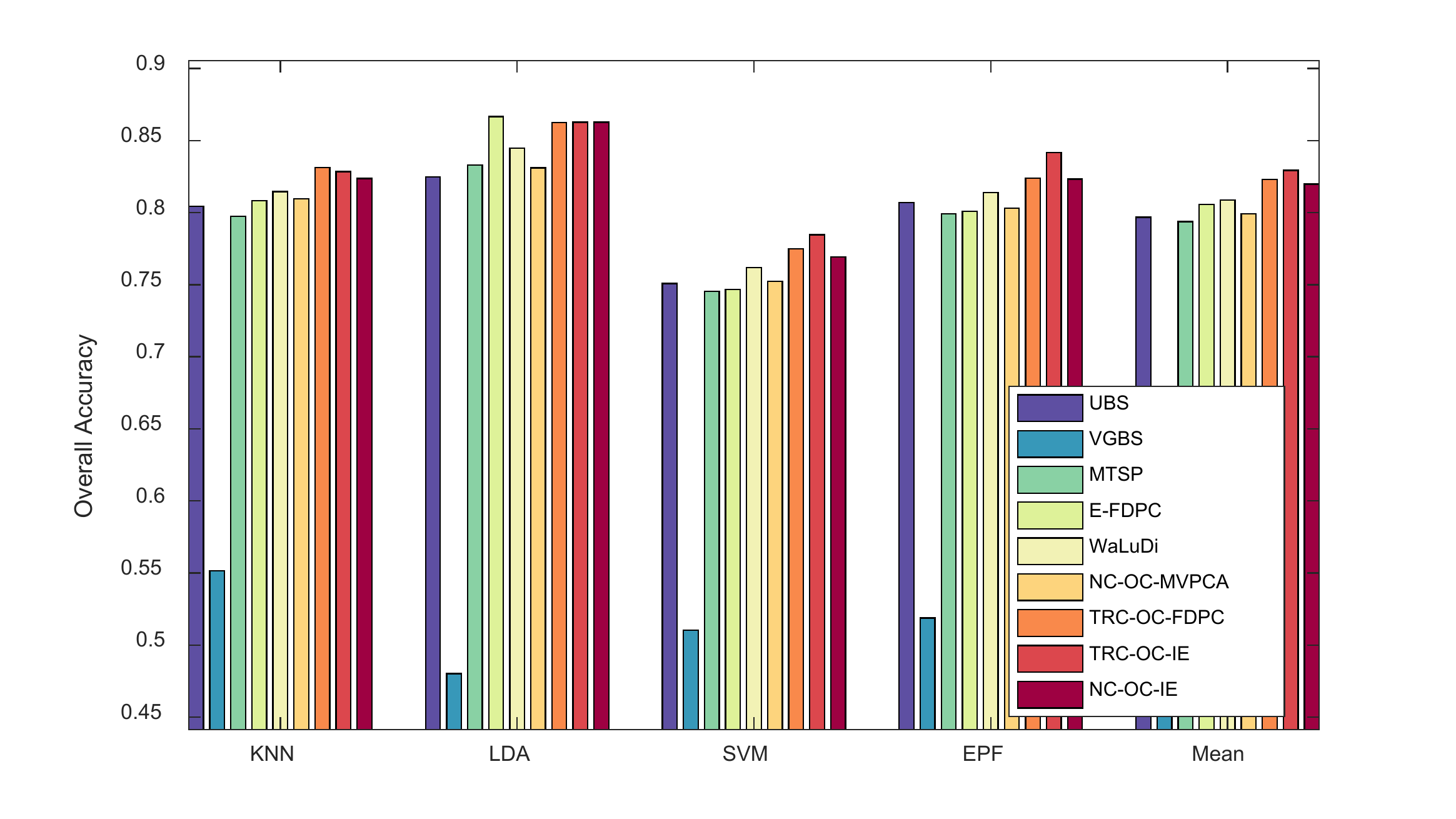,width=8.0cm}}
\end{minipage}
\caption{Average OA values against the first $30$ bands on KSC for different band selection methods.}
\label{fig:KSC_MeanOA}
\end{figure}

\begin{figure}[htb]
\begin{minipage}[b]{1.0\linewidth}
  \centering
  \centerline{\epsfig{figure=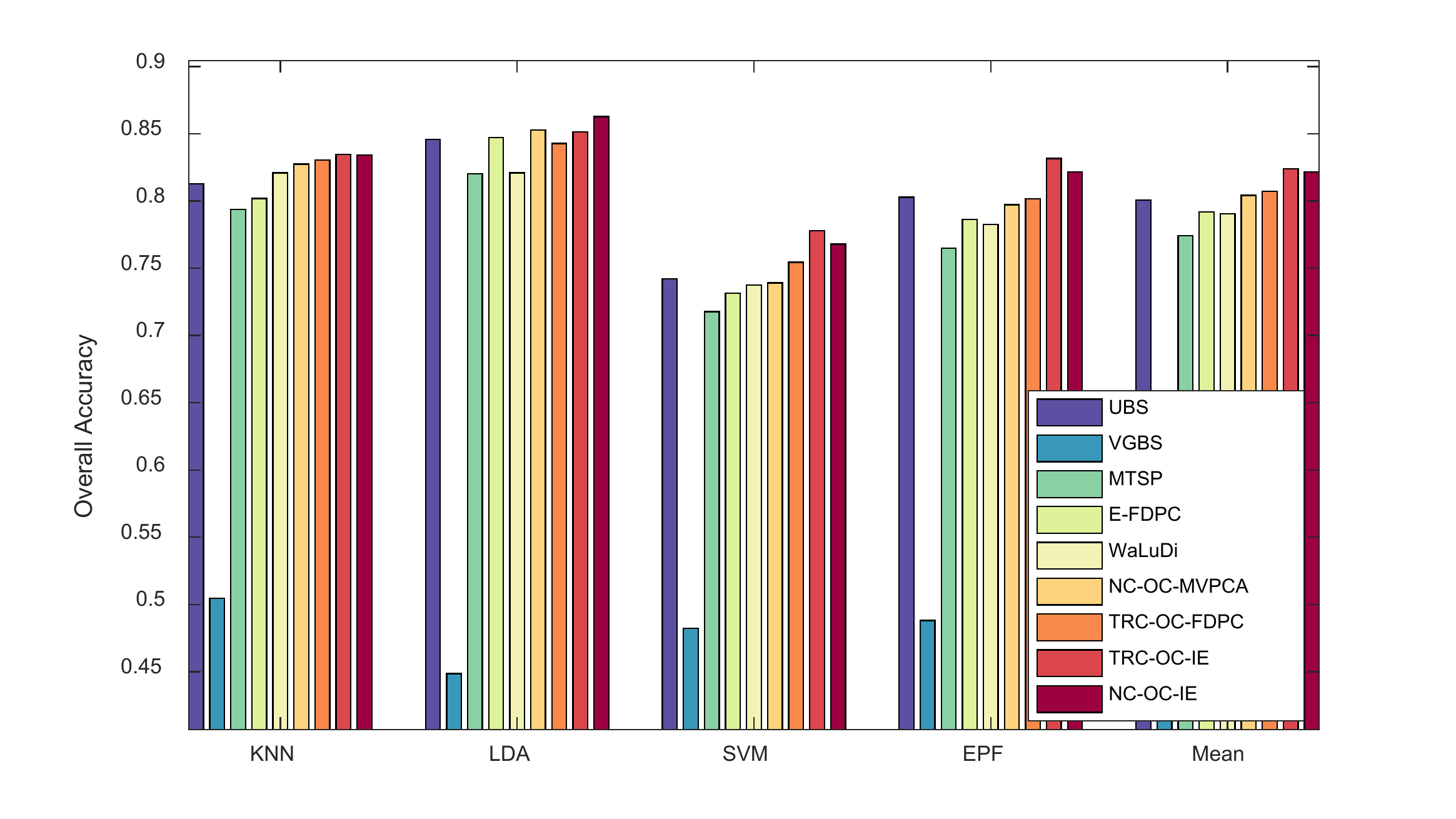,width=8.0cm}}
\end{minipage}
\caption{OA values on KSC for different band selection methods when $10$ bands are selected.}
\label{fig:KSC_OANumber}
\end{figure}

\textbf{Kennedy Space Center.}
Similar to the above data sets, Fig. \ref{fig:KSC_OACurves}, \ref{fig:KSC_MeanOA} and \ref{fig:KSC_OANumber} show the results of KSC.
In Fig. \ref{fig:KSC_OACurves}, one can observe that NC-OC-IE and TRC-OC-FDPC show significant superiority to the others on KNN, SVM and EPF.
While for LDA, E-FDPC also attains good performance.
In Fig. \ref{fig:KSC_MeanOA}, three versions of the proposed algorithm dominate the others.
However, NC-OC-MVPCA achieves a relatively worse result.
In Fig. \ref{fig:KSC_OANumber}, the proposed algorithms achieve excellent and robust performance, especially for TRC-OC-IE and NC-OC-IE.

In the following, we will discuss about some interesting issues and give some deep analyses towards the experimental results.

1) Stability against classifiers.
From the results, some competitors have instable performance on different classifiers.
For example, VGBS performs the best on Salinas Scene when EPF is employed,
but rather poor when KNN or LDA are utilized.
Similarly, E-FDPC is superior to the others on Pavia University when KNN is used,
but not for the other classifiers, especially when more than $15$ bands are selected.
This may be blamed on the selection of noisy bands,
which results in poor performance on classifiers that are sensitive to noises.
Compared to the other competitors,
the overall performance of the proposed algorithms is more robust against classifiers.
This proves that they are noise-insensitive to some extent.

2) Robustness against data sets.
Some competitors are also not robust enough against different data sets.
For instance, E-FDPC achieves high OA values on Salinas Scene,
but ranks the last on Indian Pines Scene.
This is because it is an extremely complex problem to select the optimal bands in HSI,
and local optimal solution, which will cause instability is always unavoidable.
Moreover, if there are parameters to be tuned for a method,
fluctuation in accuracy may also occur.
Fortunately, these problems can be well solved by OCF,
since it provides a non-parameter way to achieve the optimal clustering structure under CBIC.

3) Performance of UBS.
It is surprising that UBS achieves a good performance
through a simple strategy that to select the bands uniformly.
Here we try to give an explanation from physical point of view.
As stated in Section \ref{sec:CBIC}, bands with close wavelengths usually have stronger correlation.
In UBS, the minimum difference of wavelength between two adjacent bands is maximized,
so the correlation among the selected bands is reduced.
From another point of view,
the good performance of UBS proves that difference of wavelength is a good measurement for band correlation.
Hence imposing the proposed CBIC on band clustering is suitable for HSI data sets.

\begin{figure}[htb]
\begin{minipage}[b]{1.0\linewidth}
  \centering
  \centerline{\epsfig{figure=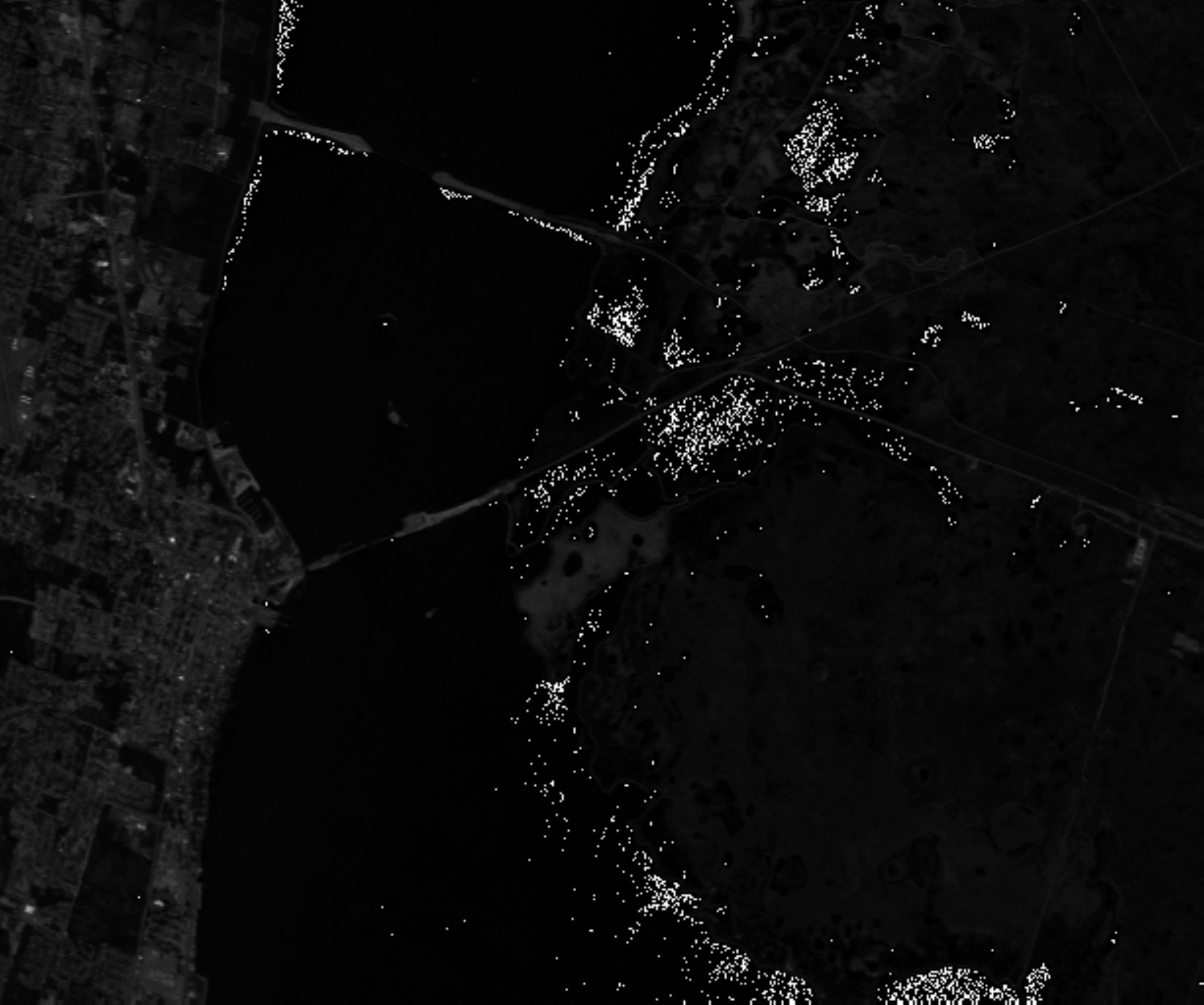,width=6.0cm}}
\end{minipage}
\caption{Band 170 of KSC data set, from which lots of salt noises can be seen. In fact, this problem also occurs in many other bands.}
\label{fig:KSC_noises}
\end{figure}
4) Performance on KSC data sets.
In fact, KSC data set is seriously polluted by salt noises (pixels with abnormally high intensities) as shown in Fig. \ref{fig:KSC_noises}.
Hence, the performance on KSC data sets can examine whether a method is robust to noises.
According to Fig. \ref{fig:KSC_MeanOA}, NC-OC-MVPCA and VGBS both achieve relatively poor performance.
This is in accordance with the analyses in Section \ref{sec:related_work} that
the variance-based MVPCA and geometry-based VGBS are very sensitive to noises.
On the other hand, the other versions of the proposed algorithm achieve satisfactory results on KSC.
This further proves that they are noise-insensitive.

5) Band number determination.
According to the plotted OA curves,
Eq. (\ref{Eq:determine_number}) offers a promising estimation of $K$.
One can observe this from two aspects.
First, as illustrated in Fig. \ref{fig:Indian_Pines_OACurves}, \ref{fig:Pavia_University_OACurves}, \ref{fig:Salinas_OACurves} and \ref{fig:KSC_OACurves},
the increasing rate of OA values tends to be slower at band numbers around the estimations.
Second, the relative relation of the required band numbers among different data sets is captured.
For instance, Pavia University has a slower increasing rate than Salinas
as shown in Fig. \ref{fig:Pavia_University_OACurves} and \ref{fig:Salinas_OACurves}.
Accordingly, the required band number of Pavia University is $17$, more than that of Salinas ($11$).
However, this method still has room for improvement.
On KSC data set, the estimation of band number is less than the real value.
This is because too large band power is assigned to the noisy bands by MVPCA.



\subsection{Comparison of Computational Time}
In order to evaluate the efficiency of the proposed algorithms,
we compare the computational times of various methods on the above four HSI data sets.
The experiments are conducted using an Intel Core i5-4590 3.30-GHz CPU with 16-GB RAM, and all the methods are implemented in MATLAB R2016b.
Table \ref{table:temporal_cost} shows the processing time of different band selection methods when selecting 15 bands on each data set.
According to it, we can find that all versions of the proposed algorithm cost moderate computational time among the other methods.
Though they are not as efficient as VGBS and E-FDPC, they take less time compared to MTSP and WaLuDi.
Therefore, the proposed algorithms are computationally acceptable while guaranteeing the superior performance.

\begin{table*}[htbp] \small
    \centering
        \caption{Processing time of different band selection methods to select 15 bands on different data sets}
            \begin{tabular}{p{0.9in}<{\centering}  p{0.9in}<{\centering}  p{0.9in}<{\centering}  p{0.9in}<{\centering}  p{0.9in}<{\centering} }
                \toprule
                           & Indian Pines & Pavia University  & Salinas & KSC\\ \midrule
                NC-OC-MVPCA & 0.40s & 0.85s & 1.08s & 2.48s\\
                NC-OC-IE & 0.50s & 0.77s & 0.91s & 1.84s\\
                TRC-OC-IE & 0.48s & 0.81s & 1.03s & 1.91s\\
                TRC-OC-FDPC & 0.51s & 0.72s & 1.15s & 1.98s\\
                VGBS & 0.21s & 0.21s & 0.37s & 0.76s\\
                MTSP & 16.59s & 9.52s & 17.26s & 17.39s\\
                E-FDPC & 0.07s & 0.24s & 0.29s & 0.79s\\
                WaLuDi & 1.69s & 7.66s & 10.20s & 36.14s\\
                \bottomrule
            \end{tabular}
\label{table:temporal_cost}
\end{table*}

\section{Conclusion}
\label{sec:conclusion}
In this paper, an optimal clustering framework (OCF) is proposed
to search for the optimal clustering structure on HSI,
and a rank on clusters strategy (RCS) is developed to select the discriminative bands under the achieved clustering structure.
A correlation-reduced band-power ratio is also presented to automatically determine the number of the required bands.
Based on the proposed OCF and RCS,
several versions of algorithm are devised by applying various objective functions and ranking methods.
Experiments on four data sets demonstrate that they are robust and effective.

In the future, we will focus on two issues to improve the proposed framework.
One is to learn the low-dimensional manifold embedded in HSI,
so as to construct a more effective similarity matrix to describe the correlation among bands.
Another is to design a more powerful objective function to capture the discrimination of band subsets.


%

%
%
%
%
%
%
%
%



\small
\bibliographystyle{IEEEtran}
\bibliography{strings}

\begin{thebibliography}{10}
\providecommand{\url}[1]{#1}
\csname url@samestyle\endcsname
\providecommand{\newblock}{\relax}
\providecommand{\bibinfo}[2]{#2}
\providecommand{\BIBentrySTDinterwordspacing}{\spaceskip=0pt\relax}
\providecommand{\BIBentryALTinterwordstretchfactor}{4}
\providecommand{\BIBentryALTinterwordspacing}{\spaceskip=\fontdimen2\font plus
\BIBentryALTinterwordstretchfactor\fontdimen3\font minus
  \fontdimen4\font\relax}
\providecommand{\BIBforeignlanguage}[2]{{%
\expandafter\ifx\csname l@#1\endcsname\relax
\typeout{** WARNING: IEEEtran.bst: No hyphenation pattern has been}%
\typeout{** loaded for the language `#1'. Using the pattern for}%
\typeout{** the default language instead.}%
\else
\language=\csname l@#1\endcsname
\fi
#2}}
\providecommand{\BIBdecl}{\relax}
\BIBdecl

\bibitem{DBLP:journals/tgrs/LuoYCZ13}
B.~Luo, C.~Yang, J.~Chanussot, and L.~Zhang, ``Crop yield estimation based on
  unsupervised linear unmixing of multidate hyperspectral imagery,''
  \emph{{IEEE} Trans. Geoscience and Remote Sensing}, vol.~51, no.~1, pp.
  162--173, 2013.

\bibitem{DBLP:journals/tbe/AkbariKKT10}
H.~Akbari, Y.~Kosugi, K.~Kojima, and N.~Tanaka, ``Detection and analysis of the
  intestinal ischemia using visible and invisible hyperspectral imaging,''
  \emph{{IEEE} Trans. Biomed. Engineering}, vol.~57, no.~8, pp. 2011--2017,
  2010.

\bibitem{DBLP:journals/tit/Hughes68}
G.~F. Hughes, ``On the mean accuracy of statistical pattern recognizers,''
  \emph{{IEEE} Trans. Information Theory}, vol.~14, no.~1, pp. 55--63, 1968.

\bibitem{7530874}
B.~Rasti, M.~O. Ulfarsson, and J.~R. Sveinsson, ``Hyperspectral feature
  extraction using total variation component analysis,'' \emph{IEEE
  Transactions on Geoscience and Remote Sensing}, vol.~54, no.~12, pp.
  6976--6985, 2016.

\bibitem{8116758}
B.~Liu, X.~Yu, P.~Zhang, A.~Yu, Q.~Fu, and X.~Wei, ``Supervised deep feature
  extraction for hyperspectral image classification,'' \emph{IEEE Transactions
  on Geoscience and Remote Sensing}, vol.~PP, no.~99, pp. 1--13, 2017.

\bibitem{7896571}
W.~Sun, G.~Yang, B.~Du, L.~Zhang, and L.~Zhang, ``A sparse and low-rank
  near-isometric linear embedding method for feature extraction in
  hyperspectral imagery classification,'' \emph{IEEE Transactions on Geoscience
  and Remote Sensing}, vol.~55, no.~7, pp. 4032--4046, 2017.

\bibitem{Kang2014Feature}
X.~Kang, S.~Li, and J.~A. Benediktsson, ``Feature extraction of hyperspectral
  images with image fusion and recursive filtering,'' \emph{IEEE Transactions
  on Geoscience and Remote Sensing}, vol.~52, no.~6, pp. 3742--3752, 2014.

\bibitem{Kang2014Intrinsic}
X.~Kang, S.~Li, L.~Fang, and J.~A. Benediktsson, ``Intrinsic image
  decomposition for feature extraction of hyperspectral images,'' \emph{IEEE
  Transactions on Geoscience and Remote Sensing}, vol.~53, no.~4, pp.
  2241--2253, 2014.

\bibitem{7378270}
X.~Cao, T.~Xiong, and L.~Jiao, ``Supervised band selection using local spatial
  information for hyperspectral image,'' \emph{IEEE Geoscience and Remote
  Sensing Letters}, vol.~13, no.~3, pp. 329--333, 2016.

\bibitem{7729714}
Y.~Tang, E.~Fan, C.~Yan, X.~Bai, and J.~Zhou, ``Discriminative weighted band
  selection via one-class svm for hyperspectral imagery,'' in \emph{2016 IEEE
  International Geoscience and Remote Sensing Symposium (IGARSS)}, 2016, pp.
  2765--2768.

\bibitem{7807260}
S.~Feng, Y.~Itoh, M.~Parente, and M.~F. Duarte, ``Hyperspectral band selection
  from statistical wavelet models,'' \emph{IEEE Transactions on Geoscience and
  Remote Sensing}, vol.~55, no.~4, pp. 2111--2123, 2017.

\bibitem{5530350}
H.~Yang, Q.~Du, H.~Su, and Y.~Sheng, ``An efficient method for supervised
  hyperspectral band selection,'' \emph{IEEE Geoscience and Remote Sensing
  Letters}, vol.~8, no.~1, pp. 138--142, 2011.

\bibitem{8059826}
X.~Cao, C.~Wei, J.~Han, and L.~Jiao, ``Hyperspectral band selection using
  improved classification map,'' \emph{IEEE Geoscience and Remote Sensing
  Letters}, vol.~14, no.~11, pp. 2147--2151, 2017.

\bibitem{6977960}
J.~Feng, L.~Jiao, F.~Liu, T.~Sun, and X.~Zhang, ``Mutual-information-based
  semi-supervised hyperspectral band selection with high discrimination, high
  information, and low redundancy,'' \emph{IEEE Transactions on Geoscience and
  Remote Sensing}, vol.~53, no.~5, pp. 2956--2969, 2015.

\bibitem{6738646}
Z.~Guo, X.~Bai, Z.~Zhang, and J.~Zhou, ``A hypergraph based semi-supervised
  band selection method for hyperspectral image classification,'' in \emph{2013
  IEEE International Conference on Image Processing}, 2013, pp. 3137--3141.

\bibitem{6738664}
H.~Li, Y.~Wang, J.~Duan, S.~Xiang, and C.~Pan, ``Group sparsity based
  semi-supervised band selection for hyperspectral images,'' in \emph{2013 IEEE
  International Conference on Image Processing}, 2013, pp. 3225--3229.

\bibitem{7332945}
H.~Su, B.~Yong, and Q.~Du, ``Hyperspectral band selection using improved
  firefly algorithm,'' \emph{IEEE Geoscience and Remote Sensing Letters},
  vol.~13, no.~1, pp. 68--72, 2016.

\bibitem{6853323}
C.~Sui, Y.~Tian, Y.~Xu, and Y.~Xie, ``Unsupervised band selection by
  integrating the overall accuracy and redundancy,'' \emph{IEEE Geoscience and
  Remote Sensing Letters}, vol.~12, no.~1, pp. 185--189, 2015.

\bibitem{7888954}
M.~Zhang, J.~Ma, and M.~Gong, ``Unsupervised hyperspectral band selection by
  fuzzy clustering with particle swarm optimization,'' \emph{IEEE Geoscience
  and Remote Sensing Letters}, vol.~14, no.~5, pp. 773--777, 2017.

\bibitem{7214263}
M.~Gong, M.~Zhang, and Y.~Yuan, ``Unsupervised band selection based on
  evolutionary multiobjective optimization for hyperspectral images,''
  \emph{IEEE Transactions on Geoscience and Remote Sensing}, vol.~54, no.~1,
  pp. 544--557, 2016.

\bibitem{DBLP:journals/tnn/WangLY16}
Q.~Wang, J.~Lin, and Y.~Yuan, ``Salient band selection for hyperspectral image
  classification via manifold ranking,'' \emph{{IEEE} Trans. Neural Netw.
  Learning Syst.}, vol.~27, no.~6, pp. 1279--1289, 2016.

\bibitem{7378877}
X.~Cao, B.~Wu, D.~Tao, and L.~Jiao, ``Automatic band selection using
  spatial-structure information and classifier-based clustering,'' \emph{IEEE
  Journal of Selected Topics in Applied Earth Observations and Remote Sensing},
  vol.~9, no.~9, pp. 4352--4360, 2016.

\bibitem{7166333}
G.~Zhu, Y.~Huang, J.~Lei, Z.~Bi, and F.~Xu, ``Unsupervised hyperspectral band
  selection by dominant set extraction,'' \emph{IEEE Transactions on Geoscience
  and Remote Sensing}, vol.~54, no.~1, pp. 227--239, 2016.

\bibitem{DBLP:journals/tip/YuanZL17}
Y.~Yuan, X.~Zheng, and X.~Lu, ``Discovering diverse subset for unsupervised
  hyperspectral band selection,'' \emph{{IEEE} Trans. Image Processing},
  vol.~26, no.~1, pp. 51--64, 2017.

\bibitem{957546}
B.~Cernuschi-Frias, ``A combinatorial generalization of the stirling numbers of
  the second kind,'' in \emph{ICECS 2001. 8th IEEE International Conference on
  Electronics, Circuits and Systems (Cat. No.01EX483)}, vol.~2, 2001, pp.
  593--596.

\bibitem{DBLP:journals/tgrs/ChangDSA99}
C.~Chang, Q.~Du, T.~Sun, and M.~L.~G. Althouse, ``A joint band prioritization
  and band-decorrelation approach to band selection for hyperspectral image
  classification,'' \emph{{IEEE} Trans. Geoscience and Remote Sensing},
  vol.~37, no.~6, pp. 2631--2641, 1999.

\bibitem{DBLP:journals/tgrs/ChangW06}
C.~Chang and S.~Wang, ``Constrained band selection for hyperspectral imagery,''
  \emph{{IEEE} Trans. Geoscience and Remote Sensing}, vol.~44, no.~6, pp.
  1575--1585, 2006.

\bibitem{DBLP:journals/tgrs/UsoPSG07}
A.~M. Us{\'{o}}, F.~Pla, J.~M. Sotoca, and P.~Garc{\'{\i}}a{-}Sevilla,
  ``Clustering-based hyperspectral band selection using information measures,''
  \emph{{IEEE} Trans. Geoscience and Remote Sensing}, vol.~45, no. 12-2, pp.
  4158--4171, 2007.

\bibitem{7161371}
S.~Jia, G.~Tang, J.~Zhu, and Q.~Li, ``A novel ranking-based clustering approach
  for hyperspectral band selection,'' \emph{IEEE Transactions on Geoscience and
  Remote Sensing}, vol.~54, no.~1, pp. 88--102, 2016.

\bibitem{7729628}
H.~Zhai, H.~Zhang, L.~Zhang, and P.~Li, ``Squaring weighted low-rank subspace
  clustering for hyperspectral image band selection,'' in \emph{2016 IEEE
  International Geoscience and Remote Sensing Symposium (IGARSS)}, 2016, pp.
  2434--2437.

\bibitem{7295589}
Y.~Yuan, J.~Lin, and Q.~Wang, ``Dual-clustering-based hyperspectral band
  selection by contextual analysis,'' \emph{IEEE Transactions on Geoscience and
  Remote Sensing}, vol.~54, no.~3, pp. 1431--1445, March 2016.

\bibitem{6165389}
S.~Jia, Z.~Ji, Y.~Qian, and L.~Shen, ``Unsupervised band selection for
  hyperspectral imagery classification without manual band removal,''
  \emph{IEEE Journal of Selected Topics in Applied Earth Observations and
  Remote Sensing}, vol.~5, no.~2, pp. 531--543, April 2012.

\bibitem{su2012hyperspectral}
H.~Su and Q.~Du, ``Hyperspectral band clustering and band selection for urban
  land cover classification,'' \emph{Geocarto International}, vol.~27, no.~5,
  pp. 395--411, 2012.

\bibitem{DBLP:journals/tgrs/GengSJZ14}
X.~Geng, K.~Sun, L.~Ji, and Y.~Zhao, ``A fast volume-gradient-based band
  selection method for hyperspectral image,'' \emph{{IEEE} Trans. Geoscience
  and Remote Sensing}, vol.~52, no.~11, pp. 7111--7119, 2014.

\bibitem{6783712}
H.~Su, Q.~Du, G.~Chen, and P.~Du, ``Optimized hyperspectral band selection
  using particle swarm optimization,'' \emph{IEEE Journal of Selected Topics in
  Applied Earth Observations and Remote Sensing}, vol.~7, no.~6, pp.
  2659--2670, 2014.

\bibitem{DBLP:journals/tgrs/YuanZW15}
Y.~Yuan, G.~Zhu, and Q.~Wang, ``Hyperspectral band selection by multitask
  sparsity pursuit,'' \emph{{IEEE} Trans. Geoscience and Remote Sensing},
  vol.~53, no.~2, pp. 631--644, 2015.

\bibitem{7536149}
H.~Su, Y.~Cai, and Q.~Du, ``Firefly-algorithm-inspired framework with band
  selection and extreme learning machine for hyperspectral image
  classification,'' \emph{IEEE Journal of Selected Topics in Applied Earth
  Observations and Remote Sensing}, vol.~10, no.~1, pp. 309--320, 2017.

\bibitem{DBLP:journals/tssc/JefferisF65}
R.~P. Jefferis and K.~A. Fegley, ``Application of dynamic programming to
  routing problems,'' \emph{{IEEE} Trans. Systems Science and Cybernetics},
  vol.~1, no.~1, pp. 21--26, 1965.

\bibitem{1098129}
R.~Larson, ``Dynamic programming with reduced computational requirements,''
  \emph{IEEE Transactions on Automatic Control}, vol.~10, no.~2, pp. 135--143,
  1965.

\bibitem{Zhang2018Cluster}
L.~Zhang, W.~Wei, Y.~Zhang, C.~Shen, A.~V.~D. Hengel, and Q.~Shi, ``Cluster
  sparsity field: An internal hyperspectral imagery prior for reconstruction,''
  \emph{International Journal of Computer Vision}, no.~11, pp. 1--25, 2018.

\bibitem{Wei2017Structured}
W.~Wei, L.~Zhang, C.~Tian, A.~Plaza, and Y.~Zhang, ``Structured sparse
  coding-based hyperspectral imagery denoising with intracluster filtering,''
  \emph{IEEE Transactions on Geoscience and Remote Sensing}, vol.~PP, no.~99,
  pp. 1--17, 2017.

\bibitem{1238361}
S.~X. Yu and J.~Shi, ``Multiclass spectral clustering,'' in \emph{Proceedings
  Ninth IEEE International Conference on Computer Vision}, 2003, pp. 313--319
  vol.1.

\bibitem{DBLP:journals/pami/ShiM00}
J.~Shi and J.~Malik, ``Normalized cuts and image segmentation,'' \emph{{IEEE}
  Trans. Pattern Anal. Mach. Intell.}, vol.~22, no.~8, pp. 888--905, 2000.

\bibitem{DBLP:conf/nips/Zelnik-ManorP04}
L.~Zelnik{-}Manor and P.~Perona, ``Self-tuning spectral clustering,'' in
  \emph{Advances in Neural Information Processing Systems 17 [Neural
  Information Processing Systems, {NIPS} 2004, December 13-18, 2004, Vancouver,
  British Columbia, Canada]}, 2004, pp. 1601--1608.

\bibitem{Shannon1948A}
C.~E. Shannon, \emph{A mathematical theory of communication}.\hskip 1em plus
  0.5em minus 0.4em\relax ACM, 1948.

\bibitem{DBLP:journals/tit/CoverH67}
T.~M. Cover and P.~E. Hart, ``Nearest neighbor pattern classification,''
  \emph{{IEEE} Trans. Information Theory}, vol.~13, no.~1, pp. 21--27, 1967.

\bibitem{DBLP:journals/tgrs/BandosBC09}
T.~V. Bandos, L.~Bruzzone, and G.~Camps{-}Valls, ``Classification of
  hyperspectral images with regularized linear discriminant analysis,''
  \emph{{IEEE} Trans. Geoscience and Remote Sensing}, vol.~47, no.~3, pp.
  862--873, 2009.

\bibitem{DBLP:journals/tgrs/MelganiB04}
F.~Melgani and L.~Bruzzone, ``Classification of hyperspectral remote sensing
  images with support vector machines,'' \emph{{IEEE} Trans. Geoscience and
  Remote Sensing}, vol.~42, no.~8, pp. 1778--1790, 2004.

\bibitem{6553593}
X.~Kang, S.~Li, and J.~A. Benediktsson, ``Spectral-spatial hyperspectral image
  classification with edge-preserving filtering,'' \emph{IEEE Transactions on
  Geoscience and Remote Sensing}, vol.~52, no.~5, pp. 2666--2677, 2014.

\end{thebibliography}
%
%
%

%
\begin{IEEEbiography}[{\includegraphics[width=1in,height=1.25in,clip,keepaspectratio]{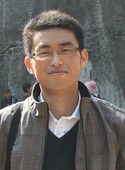}}]{Qi Wang}
(M'15-SM'15) received the B.E. degree in automation and the Ph.D. degree in pattern recognition and intelligent systems from the University of Science and Technology of China, Hefei, China, in 2005  and 2010, respectively.  He is currently a Professor with the School of Computer Science, with the Unmanned System Research Institute, and with the Center for OPTical IMagery Analysis and Learning, Northwestern Polytechnical University, Xi'an, China. His research interests include computer vision and pattern recognition.
\end{IEEEbiography}

\begin{IEEEbiography}[{\includegraphics[width=1in,height=1.25in,clip,keepaspectratio]{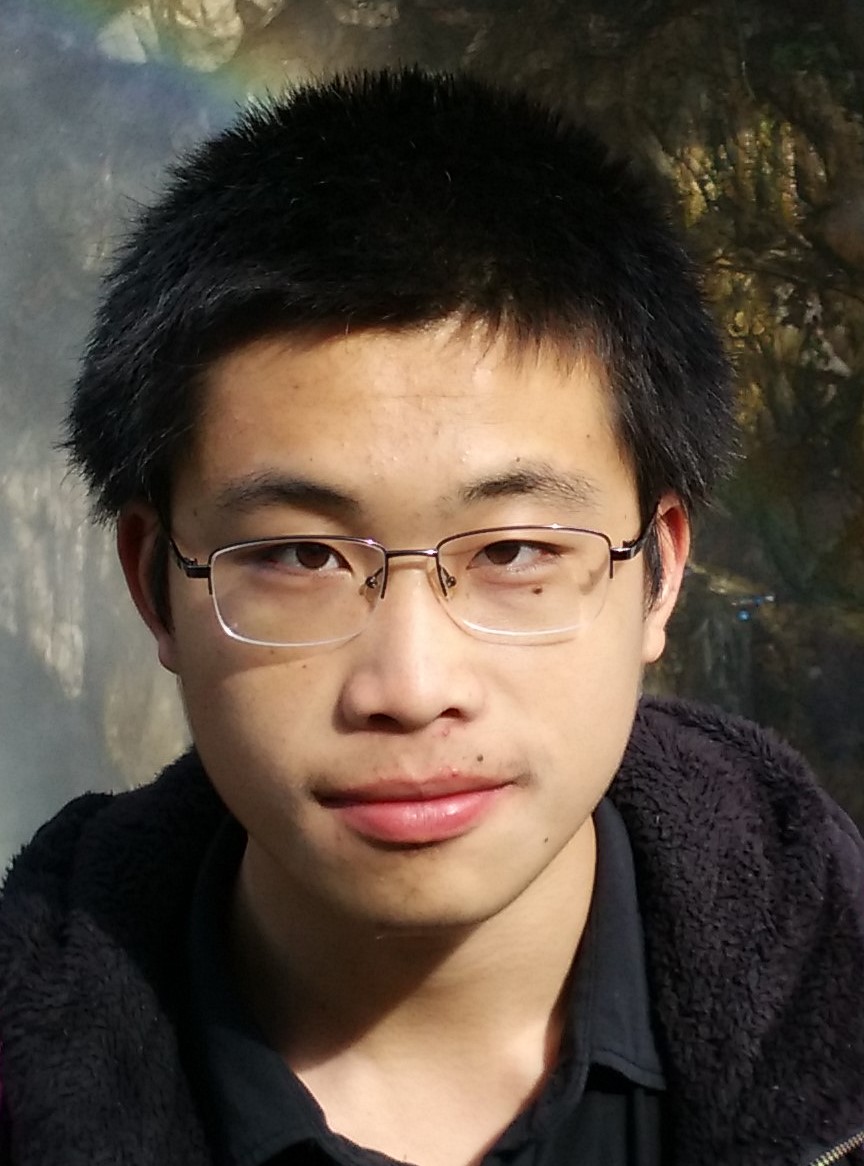}}]{Fahong Zhang}
received the B.E. degree in software engineering from the Northwestern Polytechnical University, Xi'an, China, in 2017.
He is currently working toward the M.S. degree in computer science in the Center for OPTical IMagery Analysis and Learning, Northwestern Polytechnical University, Xi'an, China.

His research interests include hyperspectral image processing and computer vision.
\end{IEEEbiography}
\begin{IEEEbiographynophoto}{Xuelong Li}
(M'02-SM'07-F'12) is a full professor with the Xi'an Institute of Optics and Precision Mechanics, Chinese Academy of Sciences, Xi'an 710119, Shaanxi, P. R. China and with the University of Chinese Academy of Sciences, Beijing 100049, P. R. China.
\end{IEEEbiographynophoto}
%
%
%




\end{document}